\documentclass[12pt]{article}
\usepackage{epsfig, amssymb}
\usepackage{graphicx,psfrag} 
\newcommand{\be}{\begin{equation}}
\newcommand{\ee}{\end{equation}}
\newcommand{\bea}{\begin{eqnarray}}
\newcommand{\eea}{\end{eqnarray}}
\newcommand{\bA}{\begin{array}}
\newcommand{\eA}{\end{array}}
\newcommand{\bc}{\begin{center}}
\newcommand{\ec}{\end{center}}
\newcommand{\al}{\alpha}

\newcommand{\ra}{\rightarrow}

\newcommand{\ie}{{\it i.e.}}
\newcommand{\eg}{{\it e.g.}}

\newcommand{\Rea}{\mathop{\rm Re}}

\newcommand{\Nt}{${\cal N}{=}2$}

\def\BC{{\mathbb C}}
\def\BP{{\mathbb P}}
\def\BR{{\mathbb R}}
\def\BZ{{\mathbb Z}}
\def\BQ{{\mathbb Q}}

\def\BN{\mbox{\boldmath$N$}}

\newcommand{\ov}{\over}

\begin{document}

\begin{titlepage}

\bc

\hfill  {TIFR/TH/06-26} \\
\hfill  {\tt hep-th/0609017} \\
         [22mm]

{\Huge Phases of unstable conifolds}
\vspace{10mm}

{\large K.~Narayan} \\
\vspace{3mm}
{\small \it Department of Theoretical Physics, \\}
{\small \it Tata Institute of Fundamental Research, \\}
{\small \it Homi Bhabha Road, Colaba, Mumbai - 400005, India.\\}
{\small Email: \ narayan@theory.tifr.res.in}\\

\ec
\medskip
\vspace{40mm}

\begin{abstract}
We explore the phase structure induced by closed string tachyon
condensation of toric nonsupersymmetric conifold-like singularities
described by an integral charge matrix $Q=(n_1\ n_2 -n_3 -n_4), \
n_i>0, \ \sum_i Q_i\neq 0$, initiated in hep-th/0510104. Using gauged
linear sigma model renormalization group flows and toric geometry
techniques, we see a cascade-like phase structure containing decays to
lower order conifold-like singularities, including in particular the
supersymmetric conifold and the $Y^{pq}$ spaces. This structure is
consistent with the Type II GSO projection obtained previously for these
singularities. Transitions between the various phases of these
geometries include flips and flops.
\end{abstract}

\end{titlepage}

\newpage 
{\small
\begin{tableofcontents}
\end{tableofcontents}
}


\section{Introduction and summary}

Understanding the stringy dynamics of nontrivial spacetime geometries
is an interesting question, especially in the absence of spacetime
supersymmetry. In this case, there typically are geometric
instabilities in the system, often stemming from closed string
tachyons in the theory (see \eg\ \cite{emilrev, minwalla0405} for
reviews), whose time dynamics is hard to unravel in detail. However
understanding the detailed phase structure of these geometries is
often tractable based on analyses of renormalization group flows in
appropriate 2-dimensional gauged linear sigma models (GLSMs)
\cite{wittenphases} describing the system with unbroken $(2,2)$ 
worldsheet supersymmetry. In this case, such an analysis closely
dovetails with the resolution of possible localized singularities
present in the space.

A simple and prototypical example of such a renormalization group flow
description of spacetime dynamics is the shrinking of a 2-sphere
($\BP^1$) given by\ $|\phi_1|^2+|\phi_2|^2=r//U(1)$. The complex
coordinates $\phi_i$ have the $U(1)$ identifications\
$(\phi_1,\phi_2)\ra (e^{i\theta}\phi_1,e^{i\theta}\phi_2)$, which we
quotient by, to obtain a 2-sphere (this symplectic quotient
construction will be elaborated on abundantly later). The parameter
$r=R^2$ is the size of the sphere. The GLSM description of this system
shows a 1-loop renormalization of the parameter $r$
\be\label{P1flow}
r=r_0+2 \log {\mu\over\Lambda}\ \qquad \equiv\qquad R^2=R_0^2-t\ .
\ee
In the equation on the right, we have recast the RG flow 
equation\footnote{This can also be obtained from studying worldsheet 
RG flow (or Ricci flow) of the 2-sphere\ 
${d\over dt}g_{\mu\nu}\sim -R_{\mu\nu}$, giving\ 
${{d\over dt}(R^2)}\sim -1$.} as an equation for the time 
evolution\footnote{Time in this paper means RG time.  Although time
evolution in spacetime is not in general the same as worldsheet RG
flow, it is consistent for the time evolution trajectories to be
qualitatively similar to the RG flow trajectories and in many known
examples, the endpoints from both approaches are identical. See \eg\ 
\cite{headrick0510, suyama} for recent related discussions: in 
particular, the worldsheet beta-function equations show that there is 
no obstruction to either RG flow (from c-theorems) or time-evolution 
(since the dilaton can be turned off) for noncompact singularities 
such as those considered here. Furthermore, for the special kinds of 
complex spaces we deal with here, the worldsheet theory has unbroken 
worldsheet supersymmetry.} of the radius by identifying the RG scale
$2\log{\mu\over\Lambda}\equiv -t$ ($\mu$ decreases along the RG flow)
and $r_0$ with the initial size $R_0^2$. Early time ($t\sim 0$ here)
corresponds to $\mu\sim\Lambda$ which in this case is $r\sim r_0\gg
0$, \ie\ large $R\sim R_0$: more generally the sign of the coefficient
of the logarithm dictates the direction of evolution of the geometry.
The RG flow shows that the sphere has an instability to shrink, with
the shrinking being slow initially since for large $R_0$, we have\
$R\sim R_0-{t\over 2R_0} +\ldots$.

This kind of behaviour also arises in the context of singular spaces
in 3 complex dimensions where much more complicated and interesting
phenomena happen.  Two types of 3-dimensional nonsupersymmetric
unstable singularities, particularly rich both in physical content and
mathematical structure, are conifolds \cite{knconiflips} and orbifolds
\cite{drmknmrp, drmkn} (see also \cite{sarkar0407}), thought of as
local singularities in some compact space, the full spacetime then
being of the form $\BR^{3,1}\times{\cal M}$. The conifold-like
singularities \cite{knconiflips} (reviewed in Sec.~2) are toric (as
are orbifolds), labelled by a charge matrix
\be
Q=(\bA{cccc} n_1 & n_2 & -n_3 & -n_4 \eA)\ , \qquad\ \ 
\qquad \sum Q_i\neq 0\ ,
\ee
for integers $n_i>0$, which characterizes their toric data 
($Q=(\bA{cccc} 1 & 1 & -1 & -1 \eA)$ corresponding to the 
supersymmetric conifold). The condition $\sum_iQ_i\neq 0$ implements 
spacetime supersymmetry breaking. It is possible to show that 
these are nonsupersymmetric orbifolds of the latter, and thus can 
be locally described by a hypersurface equation $z_1z_4-z_2z_3=0$, 
with the $z_i$ having discrete identifications from the quotienting.
Generically these spaces are not complete intersections of 
hypersurfaces. They can be described as
\be
\sum_i Q_i |\phi_i|^2 = 
n_1|\phi_1|^2 + n_2|\phi_2|^2 - n_3|\phi_3|^2 - n_4|\phi_4|^2 = r\ 
//U(1)\ ,
\ee
where the $U(1)$ gauge group acts as $\phi_i\ra e^{iQ_i\beta}\phi_i$
on the GLSM fields $\phi_i$, as will be described in detail later.
The variations of the Fayet-Iliopoulos parameter $r$ describe the
distinct phases of the geometry, with the $r\gg 0$ and $r\ll 0$
resolved phases giving fibrations over two topologically distinct
2-cycles. These $small$ $resolutions$ --- K\"ahler blowups of the
singularity (at $r=0$) by 2-cycles --- have an asymmetry stemming 
from $\sum Q_i\neq 0$. Indeed the 1-loop renormalization\
$r=(\sum_iQ_i)\log {\mu\over\Lambda}$\ shows that one of these
2-spheres $\BP^1_-$ is unstable to shrinking and the other, more
stable, $\BP^1_+$ grows. This spontaneous blowdown of a 2-cycle 
accompanied by the spontaneous blowup of a topologically distinct 
2-cycle is a flip transition. Say at early times we set up the system in the
unstable, approximately classical, (ultraviolet) phase where the
shrinking 2-sphere $\BP^1_-$ is large: then the geometry will 
dynamically evolve\footnote{Letting 
$q=-\sum_iQ_i>0,\ R_0^2=\log{\mu_0\over\Lambda}$ ($\mu_0\gg\Lambda$), 
we recast $r=q\log {\mu\over\Lambda}$ to obtain\ 
$R_-=q^{1/2}\sqrt{R_0^2-t}\sim R_0-{t\over R_0} ,\ 
R_+=q^{1/2}\sqrt{t-t_0}\sim \sqrt{t}-{t_0\over\sqrt{t}}$ for early 
($t\sim 0$) and late ($t\gg R_0^2$) times, $t_0=R_0^2$ being when
$R=0$: \ie\ the shrinking of $\BP^1_-$ and growing of $\BP^1_+$ are
slow for large $\BP^1$s.  The shrinking of $\BP^1_-$ accelerates
towards the singular region, while $\BP^1_+$ first rapidly grows, then
decelerates (within this 1-loop RG flow).} towards the more stable
$\BP^1_+$, with an inherent directionality in time, the singular
region near $r=0$ where quantum (worldsheet instanton) corrections in
the GLSM are large being a transient intermediate
state\footnote{Although one cannot make reliable statements within
this approximation about the singular region, arising as it does in
the ``middle'' of the RG flow, it is worth making a comment about the
geometry of this region. It was shown in \cite{knconiflips} (see also 
Sec.~2) that the structure of these spaces as quotients of the
supersymmetric conifold obstructs the only 3-cycle (complex structure)
deformation of the latter (although there can exist new abstract
deformations that have no interpretation ``upstairs''). This suggests
that there are no analogs of ``strong'' topology change and conifold 
transitions with nonperturbative light wrapped brane states 
here (see also the discussion on the GLSM before Sec.~3.1).}.

\begin{figure}
\bc
\epsfig{file=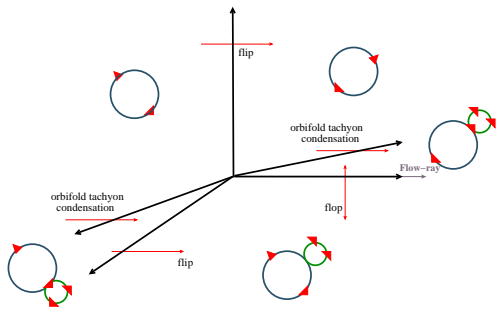, width=6.5cm}
\caption{{\small A heuristic picture of the phases of a 2-parameter system. 
The blue and green circles are $\BP^1$s and weighted $\BP^2$s respectively.
The red triangles are residual orbifold singularities on their loci.}}
\label{phases}
\ec
\end{figure}

An obvious question that arises on this analysis of \cite{knconiflips}
on the small resolutions is: 
\emph{are there RG evolution trajectories of a given unstable 
conifold-like singularity where the endpoints include the supersymmetric 
conifold, and more general lower order conifold-like singularities?}\\
In this paper, we answer this question in the affirmative. Unlike the
simple $\BP^1$ example described in (\ref{P1flow}), there typically
are orbifold singularities present on the $\BP^1_{\pm}$ loci (as
described in \cite{knconiflips}), which are themselves unstable to
resolving themselves, typically by blowups of 4-cycles (divisors)
which can be interpreted as twisted sector tachyon states in the
corresponding orbifold conformal field theories. For a large 2-sphere
$\BP^1_-$, the localized orbifold singularities on its locus are
widely separated spatially. As this $\BP^1_-$ shrinks, these pieces of
spacetime potentially containing residual singularities come together,
interact and recombine giving new spaces of distinct topology. The
existence of both 2-cycle and various 4-cycle blowup modes of the
conifold singularity besides those leading to the small resolutions
makes the full phase structure given by the GLSM quite rich. This
GLSM (also admitting $(2,2)$ worldsheet supersymmetry) with a 
$U(1)^{n+1}$ gauge group, for say $n$ additional 4-cycle blowup modes,
is described by an enlarged charge matrix $Q_i^a,\ a=1,\ldots,n+1$, 
with $n+1$ Fayet-Iliopoulos parameters $r_a$ controlling the vacuum 
structure, their RG flows describing the various phase transitions 
occurring in these geometries (a heuristic picture of the phase 
structure of a 2-parameter system is shown in Figure~\ref{phases}).
The geometry of the typical GLSM phase consists of combinations of
2-cycles and 4-cycles expanding/contracting in time, separating pieces
of spacetime described by appropriate collections of coordinate charts
glued together on their overlaps in accordance with the corresponding
toric resolution (see Figures~\ref{fig17519},~\ref{fig1746}). Besides 
flips and blowups of residual orbifold twisted sector tachyons, generic
transitions between the various distinct phases include flops (marginal 
blowdowns/blowups of 2-cycles) -- these arise along infrared moduli 
spaces. In such a case, the geometry can end up anywhere on this moduli 
space, including occasionally at (real) codimension-2 singularities on 
it: these correspond to lower order supersymmetric conifold-like spaces, 
\eg\ the $Y^{pq}$ and $L^{a,b,c}$ spaces (see Sec.3).

As discussed in \cite{knconiflips}, the GLSM RG flow for a flip
transition in fact always drives it in the direction of the (partial)
resolution leading to a less singular residual geometry, \ie\ a more
stable endpoint. This enables a classification of the phases of the
enlarged GLSMs discussed here corresponding to these unstable
singularities into ``stable'' and ``unstable'' basins of attraction,
noting the directionality of the RG trajectories involving potential
flips, which always flow towards the more stable phases. The eventual
stable phases typically consist of the stable 2-sphere $\BP^1_+$
expanding in time, alongwith the various other expanding 4-cycles
corresponding to the condensation of possible tachyons localized on
the orbifold singularities on its locus: these phases include the
various small resolutions of possible lower order conifold-like
singularities. Since the GLSM with $(2,2)$ worldsheet supersymmetry
has a smooth RG flow, the various phase transitions occurring in the
evolution of the geometry are smooth.

A nontrivial GSO projection 
\be
\sum_iQ_i=even
\ee 
was obtained in \cite{knconiflips} for the $\BR^{3,1}\times {\cal
C}^{(flip)}$ spacetime background to admit a Type II string
description with no bulk tachyons and admitting spacetime fermions. 
Here we show that the enlarged $Q_i^a$ charge matrix can be truncated
appropriately so as to obtain a phase structure consistent with this
Type II GSO projection. The final decay endpoints in Type II string 
theories are supersymmetric.

It is worth comparing these geometries to other simpler ones, \eg\
$\BC^3/\BZ_N$ orbifold singularities \cite{drmknmrp, drmkn}. In the
latter, the unstable blowup modes can be mapped explicitly to
localized closed string tachyon states arising in the twisted sectors
of the conformal field theories describing these orbifolds. A flip
transition arises when a more dominant tachyon (more negative
spacetime mass) condenses during the condensation of some tachyon,
thus corresponding to a more relevant operator in the GLSM turning on
during the RG flow induced by some relevant operator. Therefore a
careful analysis of the closed string spectrum of the orbifold
conformal field theory is in principle sufficient to understand the
decay structure of the singularity.  Generically such unstable
orbifolds decay in a cascade-like fashion to lower order orbifold
singularities which might themselves be unstable, and so on.
In the present context of the conifold-like spaces, such a conformal 
field theory description is not easy to obtain in the vicinity of the 
singular region (which arises in the ``middle'' of the RG flows, 
unlike the orbifold cases). However since the conifold transition 
itself appears to be obstructed \cite{knconiflips} (see footnote 4), 
it would seem that one could in principle use worldsheet techniques in
the early time semiclassical regions to predict the full evolution
structure. In this regard, the geometry/GLSM methods used here, aided
by the structure of the residual orbifold singularities\footnote{The
structure of nonsupersymmetric 3-dimensional orbifold singularities
\cite{drmknmrp, drmkn} is reviewed in Appendix A.} that arise in the
small resolutions, are especially powerful in obtaining an explicit
analysis. The GLSM description, dovetailing beautifully with the toric
geometry description, gives detailed insights into the phase structure
of these singularities (see Sec.~3). We analyze in detail some
examples of singularities and exhibit a cascade-like phase structure
containing lower order conifold-like singularities, including in
particular the supersymmetric conifold and the $Y^{pq}$ spaces.

\section{Some preliminaries on tachyons, flips and conifolds}

In this section, we present some generalities on the nonsupersymmetric 
conifold-like singularities in question, largely reviewing results 
presented earlier in \cite{knconiflips}. Consider a charge matrix 
\be\label{Qgen}
Q = \left( \bA{cccc} n_1 & n_2 & -n_3 & -n_4 \eA \right)
\ee
and a $\BC^*$ action on the complex coordinates 
$\Psi_i\equiv a, b, c, d$, with this charge matrix as 
$\Psi_i\ra \lambda^{Q_i}\Psi_i, \ \lambda\in\BC^*$. 
Using the redefined coordinates 
$a^{1\ov n_1},\ b^{1\ov n_2},\ c^{1\ov n_3},\ d^{1\ov n_4}$, we 
find the invariant monomials 
\be\label{zin1n2n3n4}
z_1=a^{1\ov n_1}c^{1\ov n_3} ,\ \ \ z_2=a^{1\ov n_1}d^{1\ov n_4} ,\ \ \ 
z_3=b^{1\ov n_2}c^{1\ov n_3} ,\ \ \ z_4=b^{1\ov n_2}d^{1\ov n_4}\ ,
\ee
satisfying locally 
\be\label{flipconif}
z_1 z_4 - z_2 z_3 = 0\ ,
\ee
showing that the space is locally the supersymetric conifold. Globally 
however, the phases $e^{2\pi i/n_k}$ induced on the $z_i$ by the 
independent rotations on the underlying variables $a, b, c, d$, induce 
a quotient structure on the singularity with a discrete group $\Gamma$, 
the coordinates $z_i$ having the identifications 
\bea\label{Zn1n2n3n4}
( \bA{cccc} z_1 & z_2 & z_3 & z_4 \eA )\ &\longrightarrow^{a}&
( \bA{cccc} e^{2\pi i/n_1}z_1 & e^{2\pi i/n_1}z_2 & z_3 & z_4 \eA )\ ,
\nonumber\\
&\longrightarrow^{b}&
( \bA{cccc} z_1 & z_2 & e^{2\pi i/n_2}z_3 & e^{2\pi i/n_2}z_4 \eA )\ ,
\nonumber\\
&\longrightarrow^{c}&
( \bA{cccc} e^{2\pi i/n_3}z_1 & z_2 & e^{2\pi i/n_3}z_3 & z_4 \eA )\ ,
\\
&\longrightarrow^{d}&
( \bA{cccc} z_1 & e^{2\pi i/n_4}z_2 & z_3 & e^{2\pi i/n_4}z_4 \eA )\ .
\nonumber
\eea
Thus in general the flip conifold ${\cal C}^{(flip)}$ described by\ 
$Q=(\bA{cccc} n_1 & n_2 & -n_3 & -n_4 \eA)$ is the quotient
\be\label{CFquotient}
{\cal C}^{(flip)} = {{\cal C}\over \prod_i \BZ_{n_i}}
\ee
of the supersymmetric conifold ${\cal C}$ with the action given by 
(\ref{Zn1n2n3n4}). As a toric variety described by this holomorphic 
quotient construction, this space can be described by relations between
monomials of the variables $a,b,c,d$, invariant under the $\BC^*$ 
action. In general, such spaces are not complete intersections of 
hypersurfaces, \ie\ the number of variables minus the number of 
equations is not equal to the dimension of the space. The quotient 
structure above can be shown to obstruct the only complex structure 
deformation (locally given as $z_1z_4-z_2z_3=\epsilon$) of the
supersymmetric conifold\footnote{For example, under the symmetry 
$d\ra e^{2\pi i}d$ of the underlying geometry, the $z_i$ coordinates 
transform as in (\ref{Zn1n2n3n4}), giving a nontrivial phase 
$e^{2\pi i/n_4}$ to $z_1z_4-z_2z_3$ which is inconsistent with a 
nonzero real $\epsilon$ parameter.}: there can of course be new 
abstract (non-toric) deformations which may not allow any 
interpretation in terms of the ``upstairs'' (quotient) structure.

\begin{figure}
\bc
\epsfig{file=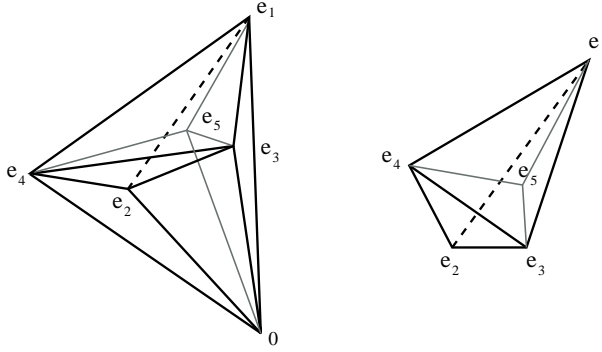, width=8cm}
\caption{{\small The toric fan for a nonsupersymmetric conifold-like 
singularity alongwith the two small resolutions $\{e_1e_2\}, 
\{e_3e_4\}$, and an interior lattice point $e_5$.}}
\label{figflip}
\ec
\end{figure}
A toric singularity corresponding to a charge matrix $Q$ can be described, 
as in Figure~\ref{figflip}, by a strongly convex rational polyhedral 
cone\footnote{A review of toric varieties and their GLSM descriptions 
appears \eg\ in \cite{morrisonplesserInstantons} (see also 
\cite{greeneCY}).} defined by four lattice vectors $e_i$ satisfying 
the relation 
\be\label{fan}
\sum Q_i e_i = n_1 e_1 + n_2 e_2 - n_3 e_3 - n_4 e_4 = 0 
\ee
in a 3-dimensional $\BN$ lattice. Assuming any three, say $e_1,e_2,e_3$, 
of the four vectors $e_i$ define a non-degenerate volume, we see using 
elementary 3-dimensional vector analysis that 
\be
(e_3-e_1)\cdot (e_2-e_1)\times(e_4-e_1) = {(\sum_i Q_i)\over n_4}\ 
e_1\cdot e_2\times e_3\ ,
\ee
so that the four lattice points $e_i$ are coplanar iff\ $\sum_i Q_i=0$. 
In this case these singularities are described as Calabi-Yau cones, 
corresponding to the $Y^{p,q}$ and $L^{a,b,c}$ spaces 
\cite{martellisparks, hanany}.

By $SL(3,\BZ)$ transformations on the lattice, one can freely choose 
two of the $e_i$, and then find the other two consistent with the 
relation (\ref{fan}). Thus fixing, say, $e_3, e_4$, we find 
\be
e_1=(-n_2,n_3k,n_4k),\qquad e_2=(n_1,n_3l,n_4l),\qquad
e_3=(0,1,0),\qquad e_4=(0,0,1)\ ,
\ee
where $k,l$ are two integers satisfying $n_1k+n_2l=1$ (assuming 
$n_1,n_2$ are coprime, $k,l$ always exist by the Euclidean algorithm).

For simplicity, we will restrict attention to the case $n_1=1$, which 
is sufficient for the physics we want to describe. In this case, we 
choose $k=1, l=0$, so that
\be\label{coneein11}
e_1=(-n_2,n_3,n_4),\qquad e_2=(1,0,0),\qquad
e_3=(0,1,0),\qquad e_4=(0,0,1)\ .
\ee
These singularities are isolated (point-like) if there are no lattice
points on the ``walls'' of the toric cone\footnote{This criterion is a
generalization of similar conditions for orbifolds \cite{drmknmrp},
reviewed in Appendix A, and for supersymmetric $Y^{p,q}, L^{a,b,c}$ 
spaces \cite{martellisparks, hanany}.}. This is true if $n_2$ is coprime 
with both of $n_3,n_4$, which can be seen as follows. If say $n_2,n_3$ 
had common factors, \ie\ say $n_2=m_1m_2, n_3=m_1m_3$ for some 
factors $m_i$, then one can construct integral lattice points\
$re_1+se_4,\ 0<r,s<1$, on the $\{e_1e_4\}$ wall: for example\footnote{We 
mention that $\{x\}=x-[x]$ denotes the fractional part of $x$, while 
$[x]$ is the integer part of $x$ (the greatest integer $\leq x$). By 
definition, $0\leq \{x\}<1$. Then for $m,n>0$, we have 
$\ [{-m\over n}]=-[{m\over n}]-1\ $ and therefore 
$\{ {-m\over n} \}=-{m\over n}-[{-m\over n}]=1-\{ {m\over n} \}$.}, 
taking $r={1\over m_1}$ and $s=1-\{ {n_4\over m_1}\}$, we have\ 
$re_1+se_4=(-m_2,m_3,{n_4\over m_1}+s)=(-m_2,m_3,[{n_4\over m_1}]+1)
\in\BN$, lying on the $\{e_1 e_4\}$ wall. Furthermore, since we can 
always write $n_4=m_4m_1+\nu$ for some $m_4$ and $\nu=0,1,\ldots,m_1-1$, 
we have\ $r+s={1\over m_1}+1-\{ {n_4\over m_1}\}<1$\ if\ 
$n_4\neq m_4m_1$ ($\nu\neq 0$), \ie\ the point $re_1+se_4$ lies 
strictly in the interior of the $\{e_1 e_4\}$ wall\ (if $n_4=m_4m_1$,
the interior point $(-m_2,m_3,m_4)={1\over m_1}e_1$ exists). 
Similarly, if $n_2,n_4$ have common factors, then there are lattice
points in the interior of the $\{e_1 e_3\}$ wall. Note that if
$n_3,n_4$ have common factors, there potentially are lattice points on
the internal $\{e_1,e_2\}$ wall.

There is a nice description of the physics of such a geometry as the
Higgs branch of the moduli space of a $U(1)$ gauged linear sigma model
admitting $(2,2)$ worldsheet supersymmetry with four scalar
superfields $\Psi\equiv \phi_1,\phi_2,\phi_3,\phi_4$, and a
Fayet-Iliopoulos (real) parameter $r$. The fields $\Psi$ transform
under $U(1)$ gauge transformations with the charge matrix $Q_i$ as
\be\label{U1gt}
\Psi_i \ra e^{i Q_i\beta} \Psi_i, \qquad\qquad Q_i = (n_1, n_2, -n_3, -n_4)\ ,
\ee
$\beta$ being the gauge parameter. The action for the GLSM is 
(using conventions of \cite{wittenphases, morrisonplesserInstantons})
\be
S = \int d^2 z\ \biggl[ d^4 \theta\ \biggl( {\bar \Psi_i} e^{2Q_i V} 
\Psi_i - {1\over 4e^2} {\bar \Sigma} \Sigma \biggr) + \Rea\biggl( i 
t\int d^2 {\tilde \theta}\ \Sigma  \biggr) \biggr]\ ,
\ee
where $t = ir + {\theta\over 2\pi}$ , $\theta$ being the $\theta$-angle 
in $1+1$-dimensions, and $e$ being the gauge coupling. The twisted 
chiral superfields $\Sigma_a$ (whose bosonic components include complex 
scalars $\sigma_a$) represent field-strengths for the gauge fields. The 
classical vacuum structure can be found from the bosonic potential 
\be\label{bospot}
U = \sum_a {(D)^2\over 2e^2} + 2{\bar\sigma}\sigma\sum_i Q_iQ_i |\Psi_i|^2\ .
\ee
Then $U=0$ requires $D=0$: solving this for $r\neq 0$ gives 
expectation values for the $\Psi_i$, which Higgs the gauge group down 
to some discrete subgroup and lead to mass terms for the $\sigma$ 
whose expectation value thus vanishes. The classical vacuum structure 
is then described by the D-term equation
\be
-{D\over e^2} = \sum_i Q_i |\Psi_i|^2 = 
n_1|\phi_1|^2 + n_2|\phi_2|^2 - n_3|\phi_3|^2 - n_4|\phi_4|^2 = r\ 
//U(1)\ ,
\ee
from which one can realize the two small resolutions (K\"ahler blowups 
by 2-cycles) as rank-2 bundles over $\BP^1_{\pm}$, as manifested by 
the GLSM moduli space for the single FI parameter ranges $r\gg 0$ 
and $r\ll 0$. These small resolutions are described in the toric fan 
by the $\{e_1,e_2\}$ and $\{e_3,e_4\}$ subdivisions: \eg\ the 
$\{e_3,e_4\}$ subdivision giving residual subcones $C(0;e_2,e_3,e_4),\ 
C(0;e_1,e_3,e_4)$, is described by the coordinate charts 
$\{(\phi_2,\phi_3,\phi_4),\ (\phi_1,\phi_3,\phi_4)\}$.
The FI parameter $r$ has a 1-loop renormalization given by
\be\label{rflow}
r = \biggl({\sum_iQ_i\over 2\pi}\biggr) \log {\mu\over \Lambda}
= \left({\Delta V\over 2\pi}\right) \log {\mu\over \Lambda}\ ,
\ee
showing that for $\sum_iQ_i\neq 0$, the GLSM RG flow drives the 
system away from the shrinking 2-sphere $\BP^1_-$, towards the 
phase corresponding to the growing 2-sphere $\BP^1_+$.\footnote{This 
has smaller $\BN$ lattice volume: the residual subcone volumes for 
the two small resolutions are\\
$\BP^1_+: V_+=V(0;e_2,e_3,e_4)+V(0;e_1,e_3,e_4)=n_1+n_2 , \ \ 
\BP^1_-: V_-=V(0;e_1,e_2,e_3)+V(0;e_1,e_2,e_4)=n_4+n_3 , 
$\\
giving the difference $\Delta V = V_+ - V_- = \sum_iQ_i$.}
This dynamical evolution process executing a flip transition 
mediates mild dynamical topology change since the blown-down 2-cycle 
$\BP^1_-$ and blown-up 2-cycle $\BP^1_+$ have distinct intersection 
numbers with various cycles in the geometry.\\
The geometric structure of the residual coordinate charts can be 
gleaned from the toric fan. From the Smith normal form algorithm of 
\cite{drmknmrp} (or otherwise), we can see that the various residual 
subcones correspond to the orbifolds\ 
$C(0;e_1,e_2,e_3)\equiv \BZ_{n_4}(1,n_2,-n_3)$, 
$C(0;e_1,e_2,e_4)\equiv \BZ_{n_3}(1,n_2,-n_4)$, and 
$C(0;e_1,e_3,e_4)\equiv \BZ_{n_2}(1,-n_3,-n_4)$, up to shifts of the 
orbifold weights by the respective orbifold orders, since these cannot 
be determined unambiguously by the Smith algorithm. Using this, one 
can see that a consistent Type II GSO projection 
\be\label{gso}
\Delta n=\sum Q_i = n_1 + n_2 - n_3 - n_4 = even
\ee
can be assigned to the conifold-like singularity in question, from the 
known Type II GSO projection $\sum k_i=even$ \cite{drmknmrp} on the 
$\BC^3/\BZ_M (k_1,k_2,k_3)$ residual orbifolds, if we make the 
reasonable assumption that a GSO projection defined for the geometry 
is not broken along the RG flows describing the decay channels.

In what follows, we will examine the phase structure of these
singularities in greater detail using their description in terms of
toric geometry and GLSMs. In particular we exhibit a cascade-like
phase structure for a singularity with given charge matrix $Q$,
containing lower order singularities $Q'$ with smaller $\sum_i Q_i'$,
consistent with the above GSO projection.

\section{The phases of unstable conifolds}

In this section, we will study the full phase structure of the unstable
conifold-like singularities in question using GLSMs and toric geometry
techniques. The prime physical observation is that the intermediate
endpoint geometries arising in the small resolution decay channels
above can contain additional blowup modes (interpreted as twisted
sector tachyons if these are residual orbifold singularities), which
further continue the evolution of the full geometry. Since these
additional blowup modes are present in the original conifold-like
singularity, there can in principle exist new decay channels
corresponding to first blowing up these modes. Technically this is
because the toric fan for such a singularity potentially contains in
its interior one or more lattice points, since the residual subcones
are potentially singular if their $\BN$ lattice volumes are greater
than unity\footnote{We recall that the $\BN$ lattice volume of an
orbifold-like cone gives the order of the orbifold singularity.}. Thus
in addition to the small resolution subdivisions \cite{knconiflips}
reviewed above, the cone $C(0;e_1,e_2,e_3,e_4)$ defining the
conifold-like singularity can also be subdivided using these interior
lattice points. In the case of orbifold singularities, the spacetime
masses of tachyons, corresponding to worldsheet R-charges of the
appropriate twisted sector operators in the orbifold conformal field
theory, effectively grade the decay channels. Since there is no such
tractable conformal field theory description for the conifold-like
geometries themselves (in the vicinity of the singularity), it is 
difficult to a priori identify their most dominant evolution channels. 
However one can efficiently resort to GLSM renormalization group
techniques (developed for unstable 3-dimensional orbifolds in
\cite{drmkn}) which essentially describe the full phase structure of
these geometries and the possible evolution patterns to the final
stable endpoints. We will first discuss the toric geometry description
and then describe some generalities of the corresponding GLSM.

Consider a singularity with charge matrix $Q$ described by the cone
defined by the $e_i,\ i=1,\ldots,4$, with one relation\
$\sum_iQ_ie_i=0$ in the 3-dimensional $\BN$ lattice. For simplicity,
we restrict attention to singularities with $n_1=1$, \ie\ of the form\
$Q=(\bA{cccc} 1 & n_2 & -n_3 & -n_4 \eA)$, with the $e_i$ given by
(\ref{coneein11}). Then as described in the previous section, there 
always exist two topologically distinct (asymmetric) small resolutions 
corresponding to the subdivisions $\{e_1e_2\}$ and $\{e_3e_4\}$: 
the subdivision $\{e_3e_4\}$ gives a less singular residual geometry 
(smaller $\BN$ lattice subcone volumes) if $n_1+n_2<n_3+n_4$. We can 
obtain detailed insight into the structure of the fan by taking
recourse to the structure of the $\BC^3/\BZ_N$ orbifold singularities
arising in these small resolution subdivisions using the techniques
and results of \cite{drmknmrp}, reviewed in Appendix A. The basic
point is that there exists a precise correspondence between operators
in the orbifold conformal field theory and $\BN$ lattice points in the
interior of (\ie\ on or below the affine hyperplane $\Delta$, described 
in Appendix A; see Figure~\ref{figorb}) the toric cone representing the 
orbifold. Thus $\BN$ lattice points in a given subcone of the toric
cone, corresponding to specific blowup modes of the singularity,
precisely map to tachyons or moduli arising in twisted sectors of the
orbifold conformal field theory corresponding to the subcone.

Now by an interior lattice point of the conifold-like cone
$C(0;e_1,e_2,e_3,e_4)$ (see Figure~\ref{figflip}), we mean lattice
points in the interior of the subcone $C(0;e_1,e_3,e_4)$ arising in
the stable small resolution (for $n_1+n_2<n_3+n_4$). Any other point
in the interior of say subcones $C(0;e_1,e_2,e_3)$ or
$C(0;e_1,e_2,e_4)$ but not $C(0;e_1,e_3,e_4)$ is effectively
equivalent to an irrelevant operator from the GLSM point of view. Now
if there exists a lattice point $e_5$ in the interior of the cone
$C(0;e_1,e_2,e_3,e_4)$, then there are two independent relations
between these five vectors $e_i,\ i=1,\ldots,5$ in the 3-dimensional
lattice $\BN$: these can be chosen as a basis for all possible
relations between these vectors. These relations
\be
\sum_i Q_i^a e_i = 0
\ee
define a charge matrix $Q_i^a$: changing the basis of 
relations amounts to changing a row of $Q_i^a$ to a rational linear 
combination of the two rows also having integral charges. Similarly,
$n$ extra lattice points in the interior of the cone give $n+1$
relations between the $e_i,\ i=1,\ldots,4+n$, thus defining a
$(n+1)\times (4+n)$ charge matrix $Q_i^a$. Specifying the structure of
this $Q_i^a$ is equivalent to giving all the information contained in
the toric fan of the singularity. For example, if there exists a
single extra lattice point $e_5$ in the interior of the subcone
$C(0;e_1,e_3,e_4)\equiv \BZ_{n_2}$, then there is a relation
of the form\ $e_5={1\over n_2} (m_1e_1 + m_3e_3 + m_4e_4),\ m_i>0$,
defining a row $Q_i^2=(\bA{ccccc} m_1 & 0 & m_3 & m_4 & -n_2 \eA)$. 
This point corresponds to a tachyon if $\sum_im_i<n_2$. Thus the 
combinatorics of $Q_i^a$ determines the geometry of the toric fan,
\eg\ whether $e_5$ is contained in the intersection of subcones say
$C(0;e_1,e_3,e_4)$ and $C(0;e_1,e_2,e_3)$, and so on.

Furthermore in Type II theories, there is a nontrivial GSO projection
that acts nontrivially on these lattice points, preserving only some
of them physically: this may be thought of as arising from the GSO
projections in the orbifold theories corresponding to the subcones
arising under the small resolutions. Thus an interior lattice point
may not in fact correspond to any blowup mode that actually exists in 
the physical theory. A simple way to encode the consequences of this 
GSO projection is to ensure that each row of the charge matrix $Q_i^a$ 
in the GLSM for the physical Type II theory sums to an even integer
\be
\sum_iQ_i^a=even\ , \qquad \ a=1,\ldots,n+1\ .
\ee 
It is easy to see that this Type II truncation of $Q_i^a$ retaining 
only rows with even sum is consistent (and we will elaborately describe 
this in examples later): \eg\ in the example above, the point 
$e_5\in C(0;e_1,e_3,e_4)$ given by 
$e_5={1\over n_2} (m_1e_1 + m_3e_3 + m_4e_4)$ defines a new 
conifold-like subcone $C(0;e_5,e_2,e_3,e_4)$, corresponding to a 
charge matrix $Q'$, which admits a Type II GSO projection iff 
$\sum_iQ_i'=even$. This constraint effectively arises from the GSO 
projection on the point $e_5$ thought of as a twisted sector state 
in the orbifold corresponding to the subcone $C(0;e_1,e_3,e_4)$.

The full phase structure of such a geometry is obtained by studying an
enlarged GLSM with gauge group $U(1)^{n+1}$ with $4+n$ superfields
$\Psi_i$ and $n+1$ Fayet-Iliopoulos parameters $r_a$. Much of the
remainder of this section is a direct generalization of the techniques
described in \cite{drmkn} to the conifold-like singularities in
question here: we present a detailed discussion primarily for 
completeness. The action of such a GLSM (in conventions of 
\cite{wittenphases, morrisonplesserInstantons}) is
\be
S = \int d^2 z\ \biggl[ d^4 \theta\ \biggl( {\bar \Psi_i} e^{2Q_i^a 
V_a} \Psi_i - {1\over 4e_a^2} {\bar \Sigma_a} \Sigma_a \biggr) + 
\Rea\biggl( i t_a\int d^2 {\tilde \theta}\ \Sigma_a  \biggr) \biggr]\ ,
\ee
where summation on the index $a=1,\ldots, n+1$ is implied. The\ $t_a = 
ir_a + {\theta_a\over 2\pi}$ \ are Fayet-Iliopoulos parameters and 
$\theta$-angles for each of the $n+1$ gauge fields ($e_a$ being the 
gauge couplings). The twisted chiral superfields $\Sigma_a$ (whose 
bosonic components are complex scalars $\sigma_a$) represent 
field-strengths for the gauge fields. The action of the $U(1)^{n+1}$ 
gauge group on the $\Psi_i$ is given in terms of the $(n+1)\times 
(4+n)$ charge matrix $Q_i^a$ above as 
\be\label{Qiagen}
\Psi_i \ra e^{i Q_i^a\lambda}\ \Psi_i\ , \qquad Q_i^a = \left( 
\bA{ccccccc} n_1 & n_2 & -n_3 & -n_4 & 0 & \ldots \\ 0 & q_2^2 & 
-q_3^2 & -q_4^2 & q_5^2 & \ldots  \\ & & \cdot & & & \ldots  \\ 
& & \cdot & & & \ldots  \\ \eA \right), \qquad \ a=1,\ldots,n+1\ .
\ee
Such a charge matrix only specifies the $U(1)^{n+1}$ action up to 
a finite group, due to the possibility of a $\BQ$-linear combination 
of the rows of the matrix also having integral charges. The specific 
form of $Q_i^a$ is chosen to conveniently illustrate specific 
geometric substructures: for example, the second row above, with 
$q_1^2=0$, describes the conifold-like subcone $C(0;e_2,e_3,e_4,e_5)$. 
The variations of the $n+1$ independent FI parameters control the 
vacuum structure of the theory. The space of classical ground states 
of this theory can be found from the bosonic potential 
\be\label{bospotgen}
U = \sum_a {(D_a)^2\over 2e_a^2} + 2\sum_{a,b} {\bar \sigma}_a 
\sigma_b \sum_i Q_i^a Q_i^b |\Psi_i|^2\ .
\ee
Then $U=0$ requires $D_a=0$: solving these for $r_a\neq 0$ gives 
expectation values for the $\Psi_i$, which Higgs the gauge group down 
to some discrete subgroup and lead to mass terms for the $\sigma_a$ 
whose expectation values thus vanish. The classical vacua of the 
theory are then given in terms of solutions to the D-term equations
\be
{-D_a\over e^2} = \sum_i Q_i^a |\Psi_i|^2 - r_a = 0\ , 
\qquad a=1,\ldots,n+1\ .
\ee
At the generic point in $r$-space, the $U(1)^{n+1}$ gauge group is 
completely Higgsed, giving collections of coordinate charts that 
characterize in general distinct toric varieties. In other words, 
this $(n+1)$-parameter system admits several ``phases'' (convex hulls 
in $r$-space, defining the secondary fan) depending on the values 
of the $r_a$. At boundaries between these phases where some (but not 
all) of the $r_a$ vanish, some of the $U(1)$s survive giving rise to
singularities classically.  Each phase is an endpoint since if left
unperturbed, the geometry can remain in the corresponding resolution 
indefinitely (within this noncompact approximation): in this sense, 
each phase is a fixed point of the GLSM RG flow. However some
of these phases are unstable while others are stable, in the sense
that fluctuations (\eg\ blowups/flips of cycles stemming from
instabilities) will cause the system to run away from the unstable 
phases towards the stable ones. This can be gleaned from the 1-loop 
renormalization of the FI parameters
\be\label{flow}
r_a = \bigg({\sum_iQ_i^a\over 2\pi}\bigg) \log {\mu\over \Lambda}\ ,
\ee
where $\mu$ is the RG scale and $\Lambda$ is a cutoff scale where 
the $r_a$ are defined to vanish. A generic linear combination of 
the gauge fields coupling to a linear combination $\sum_a\al_ar_a$ 
of the FI parameters, the $\al_a$ being arbitrary real numbers, 
has a 1-loop running whose coefficient vanishes if
\be\label{ra1loop}
\sum_{\al=1}^{n+1} \sum_{i=1}^{n+4} \al_a Q_i^a = 0\ ,
\ee
in which case the linear combination is marginal. 
This equation defines a codimension-one hyperplane perpendicular to 
a ray, called the Flow-ray, emanating from the origin and passing 
through the point $(-\sum_i Q_i^1, -\sum_i Q_i^2, \ldots, -\sum_i 
Q_i^{n+1})$ in $r$-space which has real dimension $n+1$. 
Using the redefinition\ $
{Q_i^a}'\equiv(\sum_iQ_i^1)Q_i^a-(\sum_iQ_i^a)Q_i^1 , \ a\neq 1$, 
we see that\ $\sum_i{Q_i^a}'=(\sum_iQ_i^1)(\sum_iQ_i^a)-(\sum_iQ_i^a)
(\sum_iQ_i^1)=0$, \ for $a\neq 1$, \ so that the FI parameters coupling 
to these redefined $n$ gauge fields have vanishing 1-loop running. 
Thus there is a single relevant direction (along the flow-ray) and 
an $n$-dimensional hyperplane of the $n$ marginal directions in 
$r$-space. By studying various linear combinations $\sum_a\al_ar_a$, 
we see that the 1-loop RG flows drive the system along the single 
relevant direction to the phases in the large $r$ regions of 
$r$-space, \ie, $r_a\gg 0$ (if none of the $r_a$ is marginal), 
that are adjacent to the Flow-ray \ 
$F\equiv(-\sum_iQ_i^1,-\sum_iQ_i^2,\ldots,-\sum_iQ_i^{n+1})$,\ 
or contain it in their interior: these are the stable phases.

Reversing this logic, we see that the direction precisely opposite to
the Flow-ray, \ie\
$-F\equiv(\sum_iQ_i^1,\sum_iQ_i^2,\ldots,\sum_iQ_i^{n+1})$, defines
the ultraviolet of the theory. This ray will again lie either in the
interior of some one convex hull or adjoin multiple convex hulls.
This ray $-F$ corresponds to the maximally unstable direction which 
is generically the unstable small resolution $\BP^1_-$, defining
the ultraviolet of the theory (see the examples that follow). This is
because any of the residual localized orbifold singularities on this
$\BP^1_-$ locus can be further resolved (if unstable) by turning on
the corresponding FI parameter, which process is along the Flow-ray
direction.

We restrict attention to the large $r_a$ regions, thus ignoring
worldsheet instanton corrections: this is sufficient for understanding
the phase structure, and consistent for initial values of $r_a$ whose
components in the marginal directions lie far from the center of the
marginal $n$-plane.

The 1-loop renormalization of the FI parameters can be expressed
\cite{wittenphases, wittenIAS, morrisonplesserInstantons} in terms of
a perturbatively quantum-corrected twisted chiral superpotential for
the $\Sigma_a$ for a general $n+1$-parameter system, obtained by
considering the large-$\sigma$ region in field space and integrating
out those scalars $\Psi_i$ that are massive here (and their
expectation values vanish energetically). This leads to the modified
potential
\be\label{bospotGen}
U(\sigma) = {e^2\over 2} \sum_{a=1}^{n+1} \bigg| i{\hat \tau}_a - 
{\sum_{i=1}^{4+n} Q_i^a \over 2\pi} (\log (\sqrt{2} \sum_{b=1}^{n+1} 
Q_i^b \sigma_b/\Lambda) + 1) \bigg|^2\ .
\ee
The singularities predicted classically at the locations of the phase 
boundaries arise from the existence of low-energy states at large 
$\sigma$. The physics for the nonsupersymmetric cases here is 
somewhat different from the cases where $\sum_iQ_i^a=0$ for all $a$,
as discussed in general in \cite{wittenphases, wittenIAS, 
morrisonplesserInstantons} (and for orbifold flips in \cite{drmkn}). 
Consider the vicinity of such a singularity at a phase boundary but 
far from the (fully) singular region where all $r_a$ are zero, and 
focus on the single $U(1)$ (with say charges $Q_i^1$) that is 
unbroken there (\ie\ we integrate out the other $\sigma_a,\ a\neq 1$, 
by setting them to zero). Now if $\sum_iQ_i^1=0$ (\ie\ unbroken
spacetime supersymmetry), then there is a genuine singularity when
$U(\sigma)={e^2\over 2}|i{\hat\tau}_a-{1\over 2\pi}
\sum_iQ_i^1\log|Q_i^1||^2=0$, and if $\sum_iQ_i^a=0$ for all $a$, 
this argument can be applied to all of the $U(1)$s. However for the 
nonsupersymmetric cases here, we have $\sum_iQ_i^a\neq 0$: so if say 
$\sum_iQ_i^1\neq 0$ (with the other $Q_i^a$ redefined to ${Q_i^a}'$ 
with $\sum_i{Q_i^a}'=0$), then along the single relevant direction
where $\sum_iQ_i^1\neq 0$, the potential energy has a $|\log
\sigma_1|^2$ growth. Thus the field space accessible to very low-lying
states is effectively compact (for finite worldsheet volume) and there
is no singularity for any $r_a,\theta_a$, along the RG flow: in other
words, the RG flow is smooth along the relevant direction for all
values of $\tau_1$, and the phase boundaries do not indicate
singularities.

Thus the overall physical picture is the following: the generic system
in question begins life at early times in the ultraviolet phase,
typically the unstable 2-sphere $\BP^1_-$ which has a tendency to
shrink. If this 2-sphere size is large, then this is an approximately
classical phase of the theory, with the shrinking being very slow
initially.  This $\BP^1_-$ typically has residual localized
orbifold singularities which are widely separated for a large
$\BP^1_-$. As the 2-sphere shrinks, tachyons localized at these
orbifolds might condense resolving the latter by 4-cycle blowup
modes. As the system evolves, these various cycles interact and
recombine potentially via several topology-changing flip transitions
until the geometry ultimately settles down into any of the stable
phases (which typically have distinct topology). A stable phase
typically consists of the stable 2-sphere $\BP^1_+$ growing in time,
with the various possible orbifold singularities on its locus
resolving themselves by tachyon condensation\footnote{Note that 
these conifold-like singularities always contain the small
resolutions which are K\"ahler blowup modes.  However since the Type
II GSO projection only preserves some of the K\"ahler blowup modes in
the geometry, some of the residual endpoint orbifold singularities
arising under the small resolutions could be ``string-terminal'' (as
described in \cite{drmknmrp}). In other words, these residual
orbifolds cannot be completely resolved solely by K\"ahler blowup
modes (corresponding to GSO-preserved twisted sector tachyons/moduli
in the chiral ring). Indeed since these residual orbifolds can now be
described by conformal field theory, we see the existence of
non-K\"ahler blowup modes corresponding to twisted sector tachyons
arising in any of the various (anti-)chiral rings. Thus since in the
Type II theory, there is no (all-ring) terminal $\BC^3/\BZ_N$ orbifold
singularity \cite{drmknmrp}, the final decay endpoints of the
conifold-like singularity are smooth.}. The transitions occurring 
in the course of this evolution between various phases are smooth 
as discussed above. 

In what follows, we describe two 2-parameter examples in some detail
illustrating the above generalities: one corresponds to a singularity
that has a unique late-time endpoint (within this 2-parameter
approximation), while the other includes the supersymmetric conifold
in its final endpoints, thus exhibiting infrared moduli
representing the flop between the two topologically distinct small
resolutions of the latter. Before doing so, we mention a simple 
example of a singularity which has no interior lattice point (as 
defined earlier), and evolves to its stable small resolution. The 
singularity $Q=(\bA{cccc} 1 & 1 & -1 & -3 \eA)$ is the simplest
unstable Type II conifold-like singularity. The stable small
resolution given by the subdivision $\{e_3e_4\}$ completely resolves
the singularity, since the subcone $C(0,e_1,e_3,e_4)$, potentially 
an orbifold singularity, is in fact smooth. The other small 
resolution gives rise to the orbifold subcone $C(0,e_1,e_2,e_3)\equiv 
\BZ_3 (1,1,2)$ which is effectively supersymmetric since its only 
GSO-preserved blowup mode is a marginal twisted sector state 
arising in one of the anti-chiral rings (the subcone 
$C(0,e_1,e_2,e_4)$ is smooth).

\subsection{Decays to a single stable phase}

\begin{figure}
\bc
\epsfig{file=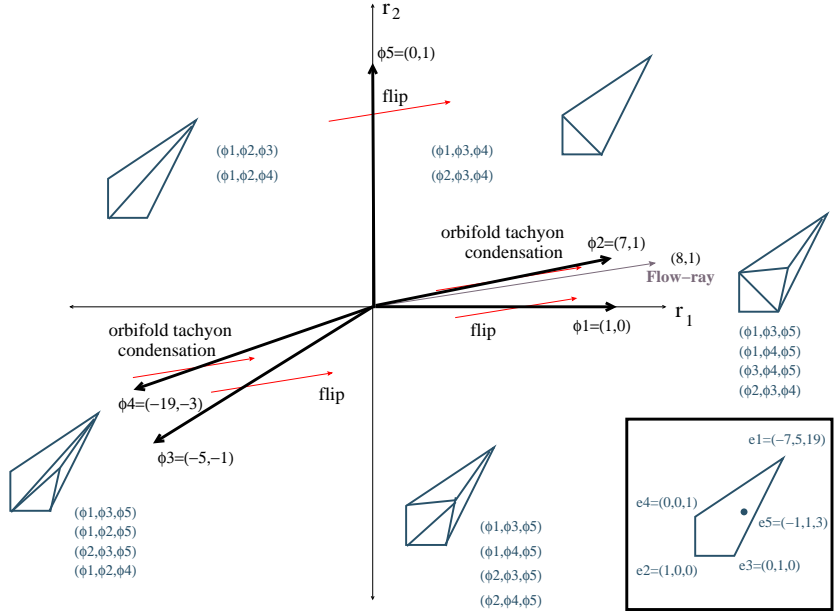, width=11cm}
\caption{{\small Phases of\ $Q=(1\ 7\ -5\ -19)$, with the toric 
subdivisions and corresponding coordinate charts in each phase, as 
well as the RG flow directions and the physics of each phase boundary.}}
\label{fig17519}
\ec
\end{figure}

Consider the singularity $Q=(\bA{cccc} 1 & 7 & -5 & -19 \eA)$\ 
(see Figure~\ref{fig17519}). The subcones can be identified as the 
following Type II orbifolds:
\be
C(0;e_1,e_2,e_3)\equiv \BZ_{19}(1,7,14)\ , \  
C(0;e_1,e_2,e_4)\equiv \BZ_5(1,2,1)\ , \ 
C(0;e_1,e_3,e_4)\equiv \BZ_7(1,2,-5)\ ,
\ee
while $C(0;e_2,e_3,e_4)$ is of course smooth. It is straightforward 
to see that 
\be
e_5\equiv (-1,1,3) = {1\over 7} (e_1 + 2e_3 + 2e_4)\ \ \in\ \ 
C(0;e_1,e_3,e_4)
\ee
corresponds to the tachyon in the twisted sector $j=1$, having 
R-charge $R_j=({1\over 7},{2\over 7},{2\over 7})$\ (GSO preserved since 
$E_j=-1$ using (\ref{gsoEj})). Including this lattice point gives the 
charge matrix 
\be\label{Qia17519}
Q_i^a = \left( \bA{ccccc} 1 & 7 & -5 & -19 & 0  \\ 0 & 1 & -1 
& -3 & 1  \\ \eA \right)\ ,
\ee
where we have used the conifold-like relation\ $e_2+e_5-e_3-3e_4=0$ to 
define the second row. Note $\sum_iQ_i^a=even,\ a=1,2$, incorporating 
the GSO projection. One could equally well have defined the second 
row in $Q_i^a$ as $(\bA{ccccc} 1 & 0 & 2 & 2 & -7 \eA)$ noticing as 
above that $e_5\in C(0;e_1,e_3,e_4)$: this does not change the physics.

To understand the phase structure of this theory, let us analyze 
the D-term equations (suppressing the gauge couplings)
\bea\label{Dterms17519}
-D_1 &=& |\phi_1|^2 + 7 |\phi_2|^2 - 5 |\phi_3|^2 - 19 |\phi_4|^2 
- r_1 = 0\ , \nonumber\\
-D_2 &=& |\phi_2|^2 + |\phi_5|^2 - |\phi_3|^2 - 3 |\phi_4|^2 - r_2 = 0\ .
\eea
There are three other auxiliary D-terms too:
\bea\label{D'17519}
-D_2' &=& -D_1+7D_2 = |\phi_1|^2 + 2 |\phi_3|^2 + 2 |\phi_4|^2 
- 7 |\phi_5|^2 - (r_1-7r_2) = 0\ , \nonumber\\
-D_3' &=& -D_1+5D_2 = |\phi_1|^2 + 2 |\phi_2|^2 - 4 |\phi_4|^2 
- 5 |\phi_5|^2 - (r_1-5r_2) = 0\ , \\
-D_4' &=& -3D_1+19D_2 = 3 |\phi_1|^2 + 2 |\phi_2|^2 + 4 |\phi_3|^2 
- 19 |\phi_5|^2 - (3r_1-19r_2) = 0\ . \nonumber
\eea
These are obtained by looking at different linear combinations of the 
two $U(1)$s that do not couple to some subset of the chiral 
superfields: \eg\ the $U(1)$s giving $D_2'$ and $D_3'$ do not couple 
to $\phi_2$ and $\phi_3$ respectively. These D-terms show that the 
five rays drawn from the origin $(0,0)$ out through the points 
$\phi_1\equiv (1,0),\ \phi_2\equiv (7,1),\ \phi_3\equiv (-5,-1),\ 
\phi_4\equiv (-19,-3),\ \phi_5\equiv (0,1)$, are phase boundaries: 
\eg\ at the boundary $(7,1)$, the $U(1)$ coupling to $r_1-7r_2$ is 
unHiggsed, signalling a classical singularity due to the existence 
of a new $\sigma$-field direction.

Before analyzing the phase structure, let us can gain some insight 
into the geometry of this singularity. In the holomorphic quotient 
construction, introduce coordinates $x_i,\ i=1,\ldots,5$, corresponding 
to the lattice points $e_i$ subject to the quotient action 
$x_i\ra\lambda^{Q_i^a}x_i$ with $Q_i^a$ given in (\ref{Qia17519}). 
Then the divisors $x_i=0,\ i=1,2,3,4$, are noncompact divisors, while 
the divisor $x_5=0$ is a compact one, whose structure can be gleaned 
as follows: the $(\BC^*)^2$ action is
\bea
g_1:\ (x_1,x_2,x_3,x_4,x_5)\ &\sim& (\lambda x_1,\lambda^7 x_2,
\lambda^{-5} x_3,\lambda^{-19} x_4,x_5)\ , \nonumber\\
g_2:\ (x_1,x_2,x_3,x_4,x_5)\ &\sim& (x_1,\lambda x_2,\lambda^{-1} x_3,
\lambda^{-3} x_4,\lambda x_5)\ ,
\eea
so that on $x_5=0$, the group element $g_1g_2^{-7}(\lambda)$ has action 
\be
(x_1,x_2,x_3,x_4,0)\ \sim\ (\lambda x_1, x_2,
\lambda^2 x_3,\lambda^2 x_4,0)\ .
\ee
When the divisor is of finite size, we expect a smooth non-degenerate 
description of the 3-dimensional space, to obtain which we must 
exclude the set $(x_1,x_3,x_4)=(0,0,0)$ \footnote{More formally, in 
the fan $\{\{e_1,e_5,e_3\}, \{e_1,e_5,e_4\}, \{e_3,e_4,e_5\}\}$, 
corresponding to the complete subdivision by $e_5$, we exclude the 
intersection of coordinate hyperplanes $x_1=x_3=x_4=0$ since 
$e_1,e_3,e_4$, are not contained in any cone of the fan.}. This 
then yields a weighted projective space $\BC\BP^2_{1,2,2}$ described 
by the coordinate chart $(x_1,x_3,x_4)$, with $x_2$ being a third 
coordinate. From the symplectic quotient point of view, we see from 
the D-term $D_2'$ that the divisor $x_5=0$, obtained by setting 
$\phi_5=0$, is
\be
\left\{|\phi_1|^2 + 2 |\phi_3|^2 + 2 |\phi_4|^2 = r_1-7r_2\right\} 
//U(1)\ ,
\ee
which is $\BC\BP^2_{1,2,2}$, with $(\phi_1,\phi_3,\phi_4)=(0,0,0)$ 
being an excluded set for nonzero K\"ahler class, \ie\ $r_1-7r_2>0$.

Now we will illustrate how the classical moduli space of the GLSM 
obtained from these D-term equations reproduces the phase diagram for 
this theory, shown in Figure~\ref{fig17519}.
In the convex hull $\{\phi_1\phi_2\}$, \ie\ $0<r_2<{1\over 7}r_1$, 
$D_2,D_2'$ imply that at least one element of each set $\phi_2,\phi_5$, 
and $\phi_1,\phi_3,\phi_4$, must acquire nonzero vacuum expectation 
values: the D-term equations do not have solutions for all of these 
simultaneously zero, which is the excluded set in this phase.
Now in the region of moduli space where $\phi_2,\phi_1$ acquire vevs, 
the light fields at low energies are $\phi_3,\phi_4,\phi_5$, which 
yield a description of the coordinate chart $(\phi_3,\phi_4,\phi_5)$. 
If $\phi_2,\phi_3$ acquire vevs, the light fields describe the chart 
$(\phi_1,\phi_4,\phi_5)$. Similarly we obtain the coordinate charts 
$(\phi_1,\phi_3,\phi_5)$ and $(\phi_2,\phi_3,\phi_4)$ if $\phi_2,\phi_4$ 
and $\phi_1,\phi_5$ acquire vevs respectively. Note that each of these 
collections of nonzero vevs are also consistent with the other D-terms 
$D_1,D_3',D_4'$. Now although one might imagine a coordinate chart 
$(\phi_1,\phi_2,\phi_4)$ from $\phi_5,\phi_3$ alone acquiring nonzero 
vevs, it is easy to see that this is not possible: for if true, 
$D_2,D_2'$ imply $|\phi_5|^2>|\phi_3|^2$ and 
$|\phi_3|^2>{7\over 2}|\phi_5|^2$, which is a contradiction. Similarly 
one sees that the possible chart $(\phi_1,\phi_2,\phi_3)$ from 
$\phi_5,\phi_4$ alone acquiring vevs is disallowed in this phase.
Thus we obtain the coordinate charts $(\phi_3,\phi_4,\phi_5)$, 
$(\phi_1,\phi_4,\phi_5)$, $(\phi_1,\phi_3,\phi_5)$ and 
$(\phi_2,\phi_3,\phi_4)$ in this phase of the GLSM.

A similar analysis of the moduli space of the GLSM can be carried out
in each of the other four phases to obtain all the possible coordinate
charts characterizing the geometry of the toric variety in that phase.

There is a simple operational method \cite{drmkn} to realize the
results of the above analysis of the D-terms for the phase boundaries
and the phases of the GLSM is the following: read off each column in
$Q_i^a$ given in (\ref{Qia17519}) as a ray drawn out from the origin
$(0,0)$ in $(r_1,r_2)$-space, representing a phase boundary. Then 
the various phases are given by the convex hulls\footnote{A 
2-dimensional convex hull is the interior of a region bounded by two 
rays emanating out from the origin such that the angle subtended by 
them is less than $\pi$.} bounded by any two of the five phase 
boundaries represented by the rays $\phi_1\equiv (1,0),\ 
\phi_2\equiv (7,1),\ \phi_3\equiv (-5,-1),\ \phi_4\equiv (-19,-3),\ 
\phi_5\equiv (0,1)$. These phase boundaries divide $r$-space into 
five phase regions, each described as a convex hull of two phase 
boundaries by several possible overlapping coordinate charts obtained 
by noting all the possible convex hulls that contain it.

The coordinate chart describing a particular convex hull, say 
$\{\phi_1,\phi_2\}$, is read off as the complementary set 
$\{\phi_3,\phi_4,\phi_5\}$. Then for instance, this convex hull is 
contained in the convex hulls $\{\phi_1, \phi_5\},\ \{\phi_2, \phi_3\}$ 
and $\{\phi_2, \phi_4\}$, so that the full set of coordinate charts 
characterizing the toric variety in the phase given by this convex hull 
$\{\phi_1, \phi_2\}$ is \ $\{\ (\phi_3,\phi_4,\phi_5),\ 
(\phi_2,\phi_3,\phi_4),\ (\phi_1,\phi_4,\phi_5),\\ (\phi_1,\phi_3,\phi_5)
\ \}$. From Figure~\ref{fig17519}, we see that this phase is the 
complete resolution corresponding to the subdivision of the toric 
cone by the small resolution $\{ e_3,e_4 \}$, followed by the lattice 
point $e_5$. Physically, the geometry of this space corresponds to the 
2-cycle $\{e_3,e_4\}$ and a 4-cycle $e_5$ blowing up simultaneously 
and expanding in time, separating the spaces described by the above 
coordinate patches (which are potentially residual orbifold 
singularities). The way these pieces of spacetime are glued together 
on the overlaps of their corresponding coordinate patches is what the 
corresponding toric subdivision in Figure~\ref{fig17519} shows.
Using the toric fan, we can glean the structure of the residual 
geometry: we see that $C(0;e_2,e_3,e_4)$ and $C(0;e_3,e_4,e_5)$ 
are both smooth, being subcones of $\BN$ lattice volume unity. 
Also we see that $C(0;e_1,e_5,e_3)\equiv \BZ_2 (-1,5,4)=\BZ_2 (1,1,0),\ 
C(0;e_1,e_5,e_4)\equiv \BZ_2 (-3,19,-4)=\BZ_2 (1,1,0)$, using the 
relations\ $e_1-5e_5+2e_2-4e_4=0$ and $3e_1-19e_5+2e_2+4e_3=0$. 
Both of these orbifolds are effectively supersymmetric $\BZ_2 (1,-1)$ 
endpoints since their anti-chiral rings contain blowup moduli. Note
also that the interior lattice point $(-4,3,11)={e_1+e_5\over 2}$ is
not GSO-preserved, and thus absent in the physical Type II theory (we
see that adding this lattice point would add a new row 
$q_i'=(\bA{cccc} 1 & 4 & -3 & -11 \eA)$ to the charge matrix, 
disallowed since $\sum_i q_i'=odd$). This is also consistent with 
the fact that this point, 
$(-4,3,11)={1\over 7}(4e_1+e_3+e_4)$, can be interpreted as a 
$j=4$ twisted sector tachyon of R-charge 
$({4\over 7},{1\over 7},{1\over 7})$ in the orbifold subcone 
$C(0;e_1,e_3,e_4)\equiv \BZ_7 (1,2,-5)$, and is GSO-projected out 
($E_j=2$ using (\ref{gsoEj})).

Similarly, using Figure~\ref{fig17519}, we recognize the other 
phases as follows.\\
The convex hull $\{\phi_2, \phi_5\}$, contained in the convex hull
$\{\phi_1, \phi_5\}$,\ yields a description of the toric variety in 
this phase in terms of the coordinate charts\ $\{(\phi_1,\phi_3,\phi_4),\ 
(\phi_2,\phi_3,\phi_4)\}$, which is the subdivision of the cone by 
the small resolution $\{e_3,e_4\}$. As we have seen, 
$C(0;e_1,e_3,e_4)\equiv \BZ_7 (1,2,-5)$, with the interior lattice 
point $e_5$ mapping to the GSO-preserved $j=1$ twisted sector tachyon 
of R-charge ${5\over 7}$.
\\
The convex hull $\{\phi_4, \phi_5\}$, contained in the convex hull
$\{\phi_3, \phi_5\}$,\ gives a description of the toric variety in 
this phase in terms of the charts\ $\{(\phi_1,\phi_2,\phi_3),\ 
(\phi_1,\phi_2,\phi_4)\}$, which is the subdivision of the cone by 
the small resolution $\{e_1,e_2\}$. This is related by a flip to 
the phase $\{\phi_2, \phi_5\}$. We see that 
$C(0;e_1,e_2,e_4)\equiv \BZ_5(1,2,1)$, while the subcone 
$C(0;e_1,e_2,e_3)\equiv \BZ_{19}(1,7,14)$ contains\ 
$e_5={1\over 19} (3e_1 + 2e_2 + 4e_3)$, corresponding to the 
GSO-preserved $j=3$ tachyon with R-charge 
$({3\over 19},{2\over 19},{4\over 19})$.
\\
The convex hull $\{\phi_3, \phi_4\}$, contained in the convex hulls 
$\{\phi_3, \phi_5\},\ \{\phi_1, \phi_4\},\ \{\phi_2, \phi_4\}$,\ 
yields a description of the toric variety in this phase in terms of 
the charts\ $\{(\phi_1,\phi_3,\phi_5),\ (\phi_1,\phi_2,\phi_5),\\ 
(\phi_2,\phi_3,\phi_5),\ (\phi_1,\phi_2,\phi_4)\}$. This is the 
subdivision of the cone by the small resolution $\{e_1,e_2\}$, followed 
by the lattice point $e_5$ which corresponds to condensation of the 
orbifold tachyon mentioned above.\\
Finally the convex hull $\{\phi_1, \phi_3\}$, contained in the convex 
hulls $\{\phi_1, \phi_4\},\ \{\phi_2, \phi_3\},\ \{\phi_2, \phi_4\}$,\ 
yields a description of the toric variety in this phase in terms of 
the charts\ $\{(\phi_1,\phi_3,\phi_5),\\ (\phi_1,\phi_4,\phi_5),\ 
(\phi_2,\phi_3,\phi_5),\ (\phi_2,\phi_4,\phi_5) \}$, which is a 
subdivision by the lattice point $e_5$ related by a flip to the 
subdivisions corresponding to either of phases\ $\{\phi_3, \phi_4\},\ 
\{\phi_1, \phi_2\}$. The subcone $C(0;e_2,e_5,e_4)$ is smooth, while 
$C(0;e_2,e_5,e_3)\equiv \BZ_3 (1,1,-1)$.

The quantum dynamics of these phases is dictated by the renormalization 
group flows in the GLSM. We remind the reader that the analysis here 
is valid only for large $r_1,r_2$, (ignoring worldsheet instanton 
corrections). The two FI parameters $r_a$ have 1-loop running given by
\be
r_1(\mu) = -{16\over 2\pi} \cdot\log {\mu\over\Lambda}\ , \qquad \qquad 
r_2(\mu) = -{2\over 2\pi} \cdot\log {\mu\over\Lambda}\ ,
\ee
so that a generic linear combination has the running
\be\label{flowray51}
\al_1 r_1 + \al_2 r_2 = -{2(8\al_1 + \al_2)\over 2\pi}\cdot\log 
{\mu\over\Lambda}\ .
\ee
The coefficient shows that this parameter is marginal if \
$8\al_1+\al_2=0$ : this describes a line perpendicular to the ray 
$(8,1)$ in $r$-space, which is the Flow-ray. Since the Flow-ray lies 
in the interior of the convex hull\ $\{\phi_1,\phi_2\}$, this is 
the $unique$ stable phase, and therefore the unique final endpoint 
geometry in this theory (within this 2-parameter system): all flow 
lines must eventually end in this phase after crossing one or more 
of the phase boundaries. The phase $\{\phi_4,\phi_5\}$, containing 
$-F\equiv (-8,-1)$, is the ultraviolet of the theory, \ie\ the early 
time phase (corresponding to the unstable small resolution $\BP^1_-$ 
with residual orbifold singularities) where all flow-lines begin. It 
is straightforward to see what crossing each of the phase boundaries 
corresponds to physically: \eg\ crossing any of $\phi_1,\ \phi_3$ or 
$\phi_5$ corresponds to topology change via a flip, while a localized 
orbifold tachyon condenses in the process of crossing either of 
$\phi_2,\ \phi_4$. This shows how the RG flow in the GLSM gives rise 
to the phase structure of the conifold-like singularity 
$Q=(\bA{cccc} 1 & 7 & -5 & -19 \eA)$. Note that the final stable 
phase is less singular than all other phases\footnote{Its total 
$\BN$ lattice subcone volume\ 
$V(0;e_1,e_5,e_3)+V(0;e_1,e_5,e_4)+V(0;e_3,e_4,e_5)+V(0;e_2,e_3,e_4)
=2+2+1+1$ is less than that for all other subdivisions, as well as 
$V_+=1+7$.}.

It is interesting to note that some of the partial decays of this
singularity exhibit two lower order conifold-like singularities, \ie\
$C(0;e_2,e_5,e_3,e_4)\equiv Q'=(\bA{cccc} 1 & 1 & -1 & -3 \eA)$ and
$C(0;e_1,e_2,e_4,e_5)\equiv Q''=(\bA{cccc} 1 & 2 & -4 & -5 \eA)$.
These are both Type II singularities having $\sum_i Q_i=even$, showing
that the decay structure is consistent with the GSO projection for
these singularities. In other words, the evolution of the geometry as
described by the GLSM RG flow does not break the GSO projection.
Since both singularities are themselves unstable, the stable phase of
the full theory also includes their stable resolutions. More generally
the various different phases in fact include distinct sets of small
resolutions of these singularities.

\subsection{Decays to the supersymmetric conifold}

\begin{figure}
\bc
\epsfig{file=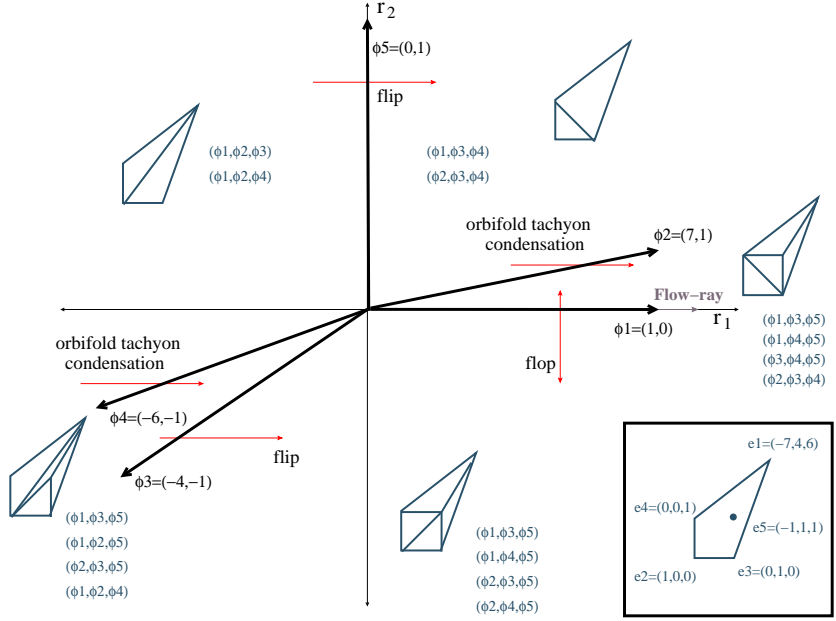, width=11cm}
\caption{{\small Phases of\ $Q=(1\ 7\ -4\ -6)$, with the toric 
subdivisions and corresponding coordinate charts in each phase, as well 
as the RG flow directions and the physics of each phase boundary.}}
\label{fig1746}
\ec
\end{figure}

Consider the singularity $Q=(\bA{cccc} 1 & 7 & -4 & -6 \eA)$ 
(see Figure~\ref{fig1746}). The various subcones arising in this fan 
can be identified as the following Type II orbifolds:
\bea
&& C(0;e_1,e_2,e_3)\equiv \BZ_6(1,1,-4)\ , \  
C(0;e_1,e_2,e_4)\equiv \BZ_4(1,-1,2)\ , \ \nonumber\\
&& C(0;e_1,e_3,e_4)\equiv \BZ_7(1,-4,1)\ , \ 
C(0;e_1,e_5,e_4)\equiv \BZ_3(1,2,1)\ ,  \  
\eea
while $C(0;e_2,e_3,e_4),\ C(0;e_1,e_5,e_3)$ are smooth. We can see that 
the lattice point
\bea
e_5\equiv (-1,1,1) &=& {1\over 7} (e_1 + 3e_3 + e_4)\ \ \in\ \ 
C(0;e_1,e_3,e_4)\ , \nonumber\\
&=& {1\over 6} (e_1 + e_2 + 2e_3)\ \ \in\ \ C(0;e_1,e_2,e_3)\ ,
\eea
corresponds to the $j=1$ twisted sector tachyon in either orbifold, 
with R-charge $R_j=({1\over 7},{3\over 7},{1\over 7})$ in 
$\BZ_7(1,-4,1)$ and $R_j=({1\over 6},{1\over 6},{1\over 3})$ in 
$\BZ_6(1,1,-4)$\ (GSO preserved since $E_j=-1$ using (\ref{gsoEj})). 
Including this lattice point gives the charge matrix 
\be\label{Qia1746}
Q_i^a = \left( \bA{ccccc} 1 & 7 & -4 & -6 & 0  \\ 0 & 1 & -1 
& -1 & 1  \\ \eA \right)\ ,
\ee
where we have used the relation\ $e_2+e_5-e_3-e_4=0$ to define the 
second row. Note $\sum_iQ_i^a=even$ for each row, consistent with 
the GSO projection. Using other relations to define $Q_i^a$ give 
equivalent physics.

The D-term equations (suppressing the gauge couplings) in this theory 
are
\bea\label{Dterms1746}
-D_1 &=& |\phi_1|^2 + 7 |\phi_2|^2 - 4 |\phi_3|^2 - 6 |\phi_4|^2 
- r_1 = 0\ , \nonumber\\
-D_2 &=& |\phi_2|^2 + |\phi_5|^2 - |\phi_3|^2 - |\phi_4|^2 - r_2 = 0\ .
\eea
The three other D-terms obtained from different linear combinations 
of the two $U(1)$s are
\bea\label{D'1746}
-D_2' &=& -D_1+7D_2 = |\phi_1|^2 + 3 |\phi_3|^2 + |\phi_4|^2 
- 7 |\phi_5|^2 - (r_1-7r_2) = 0\ , \nonumber\\
-D_3' &=& -D_1+4D_2 = |\phi_1|^2 + 3 |\phi_2|^2 - 2 |\phi_4|^2 
- 4 |\phi_5|^2 - (r_1-4r_2) = 0\ , \\
-D_4' &=& -D_1+6D_2 = |\phi_1|^2 + |\phi_2|^2 + 2 |\phi_3|^2 
- 6 |\phi_5|^2 - (r_1-6r_2) = 0\ . \nonumber
\eea
These D-terms give five phase boundaries in terms of rays drawn from 
the origin $(0,0)$ out through the points $\phi_1\equiv (1,0),\ 
\phi_2\equiv (7,1),\ \phi_3\equiv (-4,-1),\ \phi_4\equiv (-6,-1),\ 
\phi_5\equiv (0,1)$.

The phase structure of this theory, encapsulated in 
Figure~\ref{fig1746}, can be analyzed in the same way as in the 
previous case, so we will be brief here. The renormalization group 
flows in the GLSM are given by the 1-loop runnings of the two FI 
parameters $r_a$ 
\be
r_1(\mu) = -{2\over 2\pi} \cdot\log {\mu\over\Lambda}\ , \qquad \qquad 
r_2(\mu) = (0) \cdot\log {\mu\over\Lambda} + r_2^{(0)}\ .
\ee
Thus the parameter $r_2$ represents a marginal direction, and we have
explicitly shown the value $r_2^{(0)}$ of the modulus. The RG flow of 
$r_1$ however forces $r_1\ra\infty$ in the infrared. Thus the Flow-ray 
is the ray $(1,0)\equiv \phi_1$ in $r$-space (perpendicular to the
$r_2$ direction). There are two convex hulls $\{\phi_1,\phi_2\},\
\{\phi_1,\phi_3\}$, adjoining the Flow-ray, so that there are two
stable phases in this case, the $r_a$ satisfying\ 
$0<r_2<{1\over 7}r_1$ and ${1\over 4}r_1<r_2<0$ respectively. 
The ultraviolet of the theory, containing the ray $-F=(-1,0)$, is 
the phase $\{\phi_4,\phi_5\}$ corresponding to the shrinking 2-sphere 
$\BP^1_-$ with residual orbifold singularities. We can see that the 
nontrivial RG flows of the parameters $r_1-7r_2$ and $r_1-4r_2$ force 
all flowlines to cross these phase boundaries, thereby passing $into$ 
the phases $\{\phi_1,\phi_2\}$ and $\{\phi_1,\phi_3\}$ respectively. 

Physically, the geometry of, say, phase $\{\phi_1,\phi_2\}$ 
corresponds to the 2-cycle $\{e_3,e_4\}$ and the 4-cycle $e_5$ blowing 
up simultaneously and expanding in time, separating the spaces 
described by the coordinate patches $\{\ (\phi_3,\phi_4,\phi_5),\ 
(\phi_2,\phi_3,\phi_4),\ (\phi_1,\phi_4,\phi_5),\ (\phi_1,\phi_3,\phi_5)
\ \}$, with the corresponding toric subdivision in Figure~\ref{fig1746} 
showing the way these pieces of spacetime are glued together on the
overlaps of their corresponding coordinate patches.  Similarly we can
describe the geometry of the topologically distinct phase
$\{\phi_1,\phi_3\}$. The blowup mode corresponding to the 2-cycle has
size given by K\"ahler class $r_2$ which has no renormalization. This
marginality of $r_2$ physically means that in the course of the decay,
the geometry can end up anywhere on this 1-parameter moduli space. In
fact, the modulus $r_2$ corresponds to a topology-changing flop
transition interpolating between the two resolutions represented by
these phases, as can be seen from the corresponding subdivisions in
Figure~\ref{fig1746}.  Thus we expect that the geometry will sometimes
evolve precisely along the ray\
$r_2^{(0)}=r_2^c,\ r_1\ra\infty$, resulting\footnote{The classical 
singularity is at $r_2^{(0)}=0$. The constant shift 
$\tau_2^{eff}=\tau_2^{(0)}+{i\over 2\pi}\sum_iQ_i^2\log|Q_i^2|$ 
defining the singular point $r_2^{(0)}=r_2^c$, given by 
$\tau_2^{eff}=0$, arises from the bosonic potential (\ref{bospotGen}), 
since when $r_1$ is large, $\sigma_1$ is massive and can be 
integrated out (by setting $\sigma_1=0$) in (\ref{bospotGen}). 
This gives a real codimension-2 singularity after including the 
effects of the $\theta$-angle.} in the supersymmetric conifold as a 
decay product. Indeed the vevs resulting from the nonzero value of 
$r_1$ Higgs the $U(1)^2$ down to $U(1)$, thus resulting in the 
singularity (using $D_2$)
\be
\left\{|\phi_2|^2 + |\phi_5|^2 - |\phi_3|^2 - |\phi_4|^2 = 0
\right\} //U(1)\ ,
\ee
which is of course the supersymmetric conifold\ 
$Q=(\bA{cccc} 1 & 1 & -1 & -1 \eA)$. Since this is a real codimension-2 
singularity in this infrared moduli space, we expect that this is an 
occasional decay product. Generically the geometry will end up in 
either of the two stable phases $\{\phi_1,\phi_2\},\ \{\phi_1,\phi_3\}$, 
corresponding to the small resolutions (related by a flop) of this
residual singularity, obtained when $r_2>0$ and $r_2<0$ respectively,
as can be seen from the collection of coordinate charts describing the
two phases.\\
Here also, the two stable phases are less singular than any of the 
other phases.

Note that the conifold-like singularity 
$C(0;e_1,e_2,e_4,e_5)\equiv Q'=(\bA{cccc} 1 & 3 & -2 & -4 \eA)$ also 
arises among the phases of this theory: this is of course an unstable 
singularity and the flip leading to its more stable resolution connects 
the phases $\{\phi_3,\phi_4\}$ and $\{\phi_1,\phi_3\}$. This is also 
a Type II singularity, consistent with the GSO projection.

\subsection{Decays to $Y^{pq}$ spaces}

Higher order unstable singularities include, besides the supersymmetric 
conifold, the supersymmetric $Y^{pq}$ spaces defined by 
$Q=(\bA{cccc} p-q & p+q & -p & -p \eA),\ q<p$, ($p,q$ coprime), and 
$L^{a,b,c}$ spaces with $(\bA{cccc} a & b & -c & -(a+b-c) \eA), c<a+b$, 
amidst the phases arising in their evolution (see Appendix C for 
a brief description of the phase structure of the $Y^{pq}$s).

A simple subfamily of the $Y^{pq}$s is defined by 
$Q=(\bA{cccc} 1 & 2p-1 & -p & -p \eA)$. This has 
the toric cone defined by $e_5=(-(2p-1),p,p),\ e_2=(1,0,0),\ 
e_3=(0,1,0),\ e_4=(0,0,1)$. For such a singularity to arise as a 
decay product in the phases of some higher order unstable singularity, 
its cone must exist as a subcone in the cone of the latter. If we 
restrict attention to singularities of the form 
$Q=(\bA{cccc} 1 & n_2 & -n_3 & -n_4 \eA)$, then the point $e_5$ must 
be an interior point of the cone defined by $e_1=(-n_2,n_3,n_4)$ and 
$e_2,e_3,e_4$, in particular lying in the interior of the orbifold 
subcone $C(0;e_1,e_3,e_4)$. In other words, we have
\bea
&& e_5=(-(2p-1),p,p)=a(-n_2,n_3,n_4)+b(0,1,0)+c(0,0,1)\ , \nonumber\\
&& \qquad\qquad\qquad\ 0<a,b,c<1\ , \qquad  a+b+c<1\ ,
\eea
the last condition expressing $e_5$ to be a tachyon of the orbifold 
subcone $C(0;e_1,e_3,e_4)$. This then gives conditions on the $n_i$ 
\be
(p-1)n_2 < (2p-1)n_3 < pn_2\ , \qquad
(p-1)n_2 < (2p-1)n_4 < pn_2\ , \qquad 1+n_2<n_3+n_4\ .
\ee
Roughly speaking, this means that the affine hyperplane of the 
subcone $C(0;e_1,e_3,e_4)$ must be appropriately tilted so as to 
encompass the lattice point $e_5$. This gives lower bound restrictions 
on the embedding unstable singularity, the order of the embedding 
singularity rapidly rising with $p$ due to these restrictions.

For example, consider the simplest such singularity $Y^{21}\equiv 
(\bA{cccc} 1 & 3 & -2 & -2 \eA)$. Then the above conditions give 
\be
{n_2\over 3}<n_3,n_4<{2n_2\over 3}\ , \qquad 1+n_2<n_3+n_4\ ,
\ee
the first of which conditions automatically implies that the 
point $e_6=(-1,1,1)$ is also an interior point as can be checked 
by a simple calculation. This corresponds to the fact that one 
of the blowup modes of the $Y^{pq}$ singularities is the 
supersymmetric conifold (see Appendix C). 
One of the simplest unstable Type II singularities satisfying these 
conditions is $Q=(\bA{cccc} 1 & 17 & -9 & -11 \eA)$. Then we have
$C(0;e_1,e_3,e_4)\equiv \BZ_{17} (1,8,-11)$, and 
\be
e_5=(-3,2,2)={1\over 17}(3e_1+7e_3+e_4)\ , \qquad
e_6=(-1,1,1)={1\over 17}(e_1+8e_3+6e_4)
\ee 
corresponding to its GSO-preserved $j=3$ and $j=1$ twisted sector 
tachyons of R-charge $({3\over 17},{7\over 17},{1\over 17})$ and 
$({1\over 17},{8\over 17},{6\over 17})$ respectively.

Including say $e_5$ alone gives a 2-parameter system defined by 
\be
Q_i^a = \left( \bA{ccccc} 1 & 17 & -9 & -11 & 0  \\ 0 & 3 & -2
& -2 & 1  \\ \eA \right)\ ,
\ee
which can be analyzed along the same lines as before, resulting 
in the $Y^{21}$ space as an occasional decay product. Including both 
$e_5$ and $e_6$ gives a 3-parameter system with charge matrix
\be
Q_i^a = \left( \bA{cccccc} 1 & 17 & -9 & -11 & 0 & 0  \\ 0 & 3 & -2
& -2 & 1 & 0 \\ 0 & 1 & -1 & -1 & 0 & 1 \\ \eA \right)\ .
\ee
The Flow-ray for this system is $(1,0,0)\equiv \phi_1$. By analyzing
the secondary fan using the general techniques outlined earlier (and
described for a 3-tachyon system in unstable orbifolds in
\cite{drmkn}), it can be seen that there are four phases adjoining the
Flow-ray, which are the stable phases of this theory corresponding to
the various resolutions involving $Y^{21}$ and the supersymmetric
conifold contained as an interior blowup mode. It is straightforward 
to work out the details. 

More generally, these techniques show that higher order unstable
conifold singularities contain blowup modes giving rise to $L^{a,b,c}$
spaces amidst their stable phases.

\section{Discussion}

We have explored the phase structure of the nonsupersymmetric
conifold-like singularities discussed initially in \cite{knconiflips},
exhibiting a cascade-like structure containing lower order
conifold-like singularities including supersymmetric ones: this
supplements the small resolutions studied in \cite{knconiflips}. The
structure is consistent with the Type II GSO projection obtained
previously.

The GLSMs used here, as for unstable orbifolds, all have $(2,2)$
worldsheet supersymmetry, and have close connections with their
topologically twisted versions, \ie\ the corresponding A-models, so
that various physical observables (in particular those preserving
worldsheet supersymmetry) are protected along the RG flows here.
However we note that the details of the RG evolution (and therefore
also of time evolution) of the nonlinear sigma models (NLSMs)
corresponding to these conifold-like geometries can be slightly
different from the phase structure obtained here in the GLSM. For
instance, while twisted sector tachyons (and their corresponding
blowup modes) localized at the residual orbifold singularities on the
2-cycle loci have only logarithmic flows in the GLSM, on the same
footing as the 2-cycle modes, they are relevant operators in the NLSMs
with nontrivial anomalous dimensions. Thus in the NLSM (and in
spacetime), the rate of evolution of a localized tachyon mode is
expected to be higher than that of a 2-cycle mode, at least in the
large volume limit where the 2-cycle evolution is slow. However
although these details could be different, it seems reasonable, given
worldsheet supersymmetry, to conjecture that the GLSM faithfully
captures the phase structure and the evolution endpoints. A related
issue is that the marginal directions orthogonal to the flow-ray
preserved along the entire GLSM RG flow are only expected to coincide
with corresponding flat directions arising at the $final$ IR endpoints
in spacetime, which are supersymmetric as for orbifolds
\cite{drmknmrp}. However in spacetime (with broken supersymmetry), it
is not clear if there would be any corresponding exactly massless
scalar fields during the course of time evolution. Presumably this is
reconciled by taking into account the radiation effects present in
spacetime but invisible in these (dissipative) RG analyses, which may
also be related to string loop corrections (since the dilaton might be
expected to turn on).

It is worth mentioning that the classical geometry analysis in
\cite{knconiflips} on obstructions to the 3-cycle (complex structure)
deformation of these singularities due to their structure as quotients
of the supersymmetric conifold suggests that there are no analogs of
``strong'' topology change and conifold transitions with
nonperturbative light wrapped brane states here. From the GLSM point
of view, the singular region where all $r_a$ vanish arises in the
``middle'' of the RG flow and is a transient intermediate state where
the approximations in this paper are not reliable. It might be
interesting to understand the structure of instanton corrections with
a view to obtaining a deeper understanding of the physics of the
singular region encoding the flip.

On a somewhat broader note, it might be interesting to understand and
develop interconnections between renormalization group flows in
generalizations of the GLSMs considered here (and the ``space of
physical theories'' they describe) and Ricci flows in corresponding
geometric systems. The fact that the GLSM RG trajectories in the
conifold-like geometries here as well as those in \cite{drmkn} flow
towards less singular geometries (smaller $\BN$ lattice volumes)
suggests that there is a monotonically decreasing c-function-like
geometric quantity here. Physically this seems analogous to the
tachyon potential, or a height function on the ``space of
geometries''.

It would be interesting to understand D-brane dynamics in the context
of such singularities. We expect that the quivers for these D-brane
theories will be at least as rich as those for the $L^{a,b,c}$ spaces
described in \cite{hanany}, and perhaps the knowledge of the phase
structure of these theories developed here will be helpful in this
regard. It is interesting to ask what these D-brane quivers (or
possible duals) see as the manifestation of these instabilities.

Finally we make a few comments on compactifications of these
(noncompact) conifold-like singularities. We expect that such a
nonsupersymmetric conifold singularity can be embedded (classically)
in an appropriate nonsupersymmetric orbifold of a Calabi-Yau that
develops a localized supersymmetric conifold singularity, such that
the quotienting action on the latter results in the nonsupersymmetric
one. For quotient actions that are isolated, the Calabi-Yau only
acquires discrete identifications so that the resulting quotient space
``downstairs'' is locally Calabi-Yau. While we expect that the
low-lying singularities, \ie\ small $n_i$, admit such locally
supersymmetric compactifications, we note that the higher order ones
may not. In fact there may be nontrivial constraints on the $n_i$ for
the existence of such compactifications. In the noncompact case, we
note that the early time semiclassical phase is a small resolution
$\BP^1_-$ of topology distinct from that of the late time small
resolution $\BP^1_+$ phase. We expect that both these phases, being
semiclassical, admit descriptions as topologically distinct small
resolutions in compact embeddings comprising orbifolds of appropriate
Calabi-Yaus as described above. Thus one might think that the
(intermediate) flip visible explicitly in the GLSM here persists in
the compact context as well, where it would mediate mild
time-dependent topology change of the ambient compact space, with
changes in the intersection numbers of the various cycles of the
geometry.  However since in the compact context worldsheet RG
techniques are subject to the strong constraints imposed by the
c-theorem, it is not clear if our GLSM analysis here is reliable in
gaining insight into the dynamics of compact versions of the flip
transitions here (see \eg\ \cite{eva0502} for related discussions in
the context of string compactifications on Riemann surfaces). It would
be desirable to obtain a deeper understanding of these
compactifications \cite{wip} and their dynamics, perhaps implementing
the quotient action on the Calabi-Yau directly in a spacetime
description. From the latter perspective, the time dependence of the
compact internal space would imply interesting time-dependent effects
in the remaining 4-dimensional part of spacetime: for instance, in a
simple FRW-cosmology-like setup, the 4D scale factor will evolve in
accordance with the time dynamics of the internal space. It would be
interesting to explore this here perhaps along the lines of
\cite{keshav0501}.

\vspace{5mm}
{\small {\bf Acknowledgments:} I have benefitted from an early
discussion with R.~Gopakumar and from comments from S.~Minwalla 
and D.~Morrison on a draft.}


\appendix
\section{A review of $\BC^3/\BZ_N$ orbifolds: geometry and conformal 
field theory}

In this section, we review some of the features \cite{drmknmrp} of the
conformal field theory of $\BC^3/\BZ_N$ orbifold singularities, and
the way they dovetail with the toric geometry description of these
singularities. In particular, we will also review the correspondence
between operators in the orbifold conformal field theory and subspaces
in the $\BN$ lattice.

The spectrum of twisted sector string excitations in a $\BC^3/\BZ_N
(k_1,k_2,k_3)$ orbifold conformal field theory, classified using the
representations of the \Nt\ superconformal algebra, has a product-like
structure (one for each of the three complex planes) giving eight
chiral and anti-chiral rings in four conjugate pairs. A chiral ring
twist field operator has the form $X_j=\prod_{i=1}^3\ X^i_{\{jk_i/N\}}=
\prod_{i=1}^3\ \sigma_{\{jk_i/N\}}\ e^{i\{jk_i/N\}(H_i-{\bar H}_i)}$, 
where $\sigma_a$ is the bosonic twist-$a$ field operator, while the 
$H_i$ are bosonized fermions. These correspond to relevant, marginal 
and irrelevant operators with worldsheet R-charges 
$R_j\equiv (\{{jk_1\over N}\},\{{jk_2\over N}\},\{{jk_3\over N}\})
=\sum_i \{{jk_i\over N}\}$ and masses in spacetime given by 
$m_j^2={2\over\al'}(R_j-1)$. 

\begin{figure}
\bc
\epsfig{file=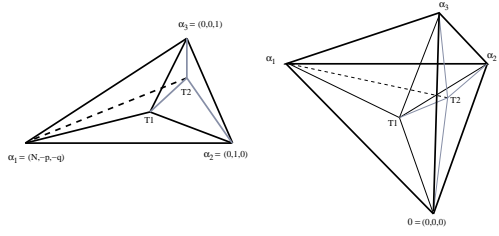, width=6.5cm}
\caption{{\small The $\BC^3/\BZ_N (1,p,q)$ orbifold toric fan, 
and tachyonic lattice points with their subdivisions.}}
\label{figorb}
\ec
\end{figure}

The geometry of such an orbifold can be recovered efficiently using 
its toric data. Let the toric cone of this orbifold be defined by 
the origin and lattice points $\al_1,\al_2,\al_3$ (see 
Figure~\ref{figorb}): the points $\al_i$ define an affine hyperplane 
$\Delta$ passing through them. The volume of this cone 
$V(0;\al_1,\al_2,\al_3)\equiv |{\rm det} (\al_1,\al_2,\al_3)|=
|\al_1\cdot \al_2\times\al_3|$ gives the order $N$ of the orbifold 
singularity\footnote{We have normalized the cone volume without any 
additional numerical factors.}. The specific structure of the orbifold 
represented by a toric cone $C(0;\al_1,\al_2,\al_3)$ can be gleaned 
either using the Smith normal form algorithm \cite{drmknmrp}, or 
equivalently by realizing relations between the lattice vectors $\al_i$ 
and any vector that is also itself contained in the $\BN$ lattice: 
\eg\ we see that the cone defined by\ $\al_1=(N,-p,-q), \al_2=(0,1,0), 
\al_3=(0,0,1)$, corresponds to $\BC^3/\BZ_N (1,p,q)$ using the 
relation $(1,0,0)={1\over N}(\al_1+p\al_2+q\al_3)$ with the lattice 
point $(1,0,0)$. Note that in general this only fixes the orbifold 
weights upto shifts by the order $N$.

There is a 1-1 correspondence between the chiral ring operators and 
points in the $\BN$ lattice toric cone of the orbifold. A given lattice 
point $P_j=(x_j,y_j,z_j)$ can be mapped to a twisted sector chiral ring
operator in the orbifold conformal field theory by realizing that this 
vector can expressed in the $ \{\al_1,\al_2,\al_3\} $ basis as
\be
(x_j,y_j,z_j) = r_1\al_1 + r_2\al_2 + r_3\al_3\ . 
\ee
If $r_i>0$, then $P_j$ is in the interior of the cone. This then 
corresponds to an operator $O_j$ with R-charge\ $R_j\equiv(r_1,r_2,r_3)$. 
Conversely, it is possible to map an operator $O_j$ of given R-charge 
to a lattice point $P_j$. There are always lattice points lying 
``above'' the affine hyperplane $\Delta$, corresponding to irrelevant 
operators: these have $R_j=\sum_ir_i>1$. Interior points lying $on$
$\Delta$ have $R_j=1$ and are marginal operators, while those ``below''
the hyperplane $\Delta$ have $R_j<1$ and correspond to 
tachyons\footnote{Note that for the $\BC^3/\BZ_N (1,p,q)$ orbifold 
(Figure~\ref{figorb}), we have the relation
\bea
{x_j(1+p+q)\over N} + y_j + z_j = r_1+r_2+r_3=R_j\ , \nonumber
\eea
so that for a supersymmetric orbifold $1+p+q=0 ({\rm mod} 2N)$, we 
have all $R_j$ integral since $x_j,y_j,z_j\in\BZ$, \ie\ there are no 
tachyonic lattice points.}. 
The toric cone of this orbifold can thus be subdivided by any of the 
tachyonic or marginal blowup modes (the irrelevant ones are unimportant 
from the physics point of view), giving rise to three residual 
subcones: these are potentially orbifold singularities again, unstable 
to tachyon condensation. For example, condensation of the tachyon 
$T=({1\over N},{p\over N},{q\over N})$ in the $\BC^3/\BZ_N (1,p,q)$ 
orbifold, corresponds to the subdivision of the cone 
$C(0;\al_1,\al_2,\al_3)$ by the interior lattice point $T\equiv(1,0,0)$. 
From the GLSM point of view, this corresponds to RG flow of the single 
Fayet-Iliopoulos parameter in a GLSM with a $U(1)$ gauge group and 
charge matrix $Q=(\bA{cccc} 1 & p & q & -N \eA)$: this gives the 
resolved phase as the stable phase. Systems of multiple tachyons in 
orbifolds can be analyzed by appropriate generalizations of this GLSM 
\cite{drmkn}, and generically exhibit flips amidst their phases.

A $\BC^3/\BZ_N (1,p,q)$ orbifold (Figure~\ref{figorb}) is isolated if 
$p,q$ are coprime w.r.t. $N$: this is equivalent to the condition 
that there are no lattice points on the walls of the defining toric 
cone. For example, if $q,N$ have a common factor $n$ with $q=m_1n,\ 
N=m_0n$, then the $\{e_1,e_2\}$ wall has the integral lattice point 
${1\over n}(N,-p,-q)+\{{p\over n}\}(0,1,0)
=(m_0,-[{p\over n}],-m_1)$. Similarly the $\{e_1,e_3\}$ wall has 
integral lattice points if $p,N$ have common factors.

There is one further important issue raised by the GSO projection 
for these residual orbifold subcones and the lattice points in their 
interior. From the results of \cite{drmknmrp}, we have that an 
orbifold $\BC^3/\BZ_N (k_1,k_2,k_3)$ admits a Type II GSO projection 
if $\sum_ik_i=even$. In addition, this GSO projection acts 
nontrivially on the twisted sector operators, preserving only some 
states in each of the four independent chiral or anti-chiral rings 
of the orbifold conformal field theory. For example, the $j$-th 
twisted sector chiral ring operator $X_j$ with R-charge 
$R_j=(\{{jk_1\over N}\},\{{jk_2\over N}\},\{{jk_3\over N}\})$ is 
GSO-preserved iff 
\be\label{gsoEj}
E_j=\sum_i \left[{jk_i\over N}\right]=odd\ .
\ee
It can be shown that under condensation of a GSO-preserved tachyon
$T_j$, the GSO projection for the residual orbifolds and residual
tachyons is consistent with this description. In other words, each of
the three residual orbifolds admits a Type II GSO projection, and
originally GSO-preserved residual tachyons continue to be
GSO-preserved after condensation of a GSO-preserved tachyon for each
of the three residual singularities.

Geometric terminal singularities arise if there is no K\"ahler blowup 
mode: \ie\ there is no relevant or marginal chiral ring operator and 
no lattice point in the interior of the toric cone. However, a physical 
analysis of the system must include all possible tachyons in all rings, 
\ie\ both K\"ahler and non-K\"ahler blowup modes. Then it turns out 
that there are no all-ring terminal singularities in Type II theories, 
while $\BC^3/\BZ_2 (1,1,1)$ is the only terminal singularity (in 
Type 0 theories). Thus the endpoint of tachyon condensation in Type II 
theories is smooth.

\section{Phase structure of $Y^{pq}$ singularities}

The $Y^{pq}$ singularities are defined by 
$Q=(\bA{cccc} p-q & p+q & -p & -p \eA)$, with $q<p$ and $p,q$ 
coprime.  More general noncompact Calabi-Yau spaces include the
$L^{a,b,c}$s which are defined by $Q=(\bA{cccc} a & b & -c & -d \eA)$,
with $\sum_iQ_i=0$. Since $\sum_iQ_i=0$ for all these, the $e_i$ 
defining the cone are coplanar, and the singularities admit a
Type II GSO projection as expected. There is no RG flow for
Fayet-Iliopoulos parameters in the corresponding GLSM and all phases
are on equal footing, defining distinct resolutions of the 
singularity.

For example, the singularity $Y^{32}$, defined by the charge matrix 
$Q=(\bA{cccc} 1 & 5 & -3 & -3 \eA)$, can be represented by the toric 
cone with $e_1=(-5,3,3),\ e_2=(1,0,0),\ e_3=(0,1,0),\ e_4=(0,0,1)$. 
There are two interior lattice points, $e_5=(-1,1,1)={e_1+2e_2\over 3}$ 
and $e_6=(-3,2,2)={2e_1+e_2\over 3}$, lying on the $\{e_1,e_2\}$ plane.
The subcones $C(0;e_5,e_2,e_3,e_4)$ and $C(0;e_6,e_2,e_3,e_4)$ 
define the lower order singularities corresponding to the 
supersymmetric conifold $Q=(\bA{cccc} 1 & 1 & -1 & -1 \eA)$ and 
$Y^{21}\equiv Q=(\bA{cccc} 1 & 3 & -2 & -2 \eA)$. 

Considering GLSMs that incorporate these interior lattice points 
gives the full phase structure of these spaces. For instance, 
including say the lattice point $e_5$ alone gives a 2-parameter GLSM 
with charge matrix
\be
Q_i^a = \left( \bA{ccccc} 1 & 5 & -3 & -3 & 0  \\ 0 & 1 & -1 
& -1 & 1  \\ \eA \right)\ ,
\ee
with two FI parameters that do not run. Since two phase boundaries 
$\phi_3=\phi_4=(-3,-1)$ coincide, we obtain four phases here instead 
of five as in the Examples in Sec.~3. We could also use the relation 
$e_5={1\over 5}(e_1+2e_3+2e_4)$ stemming from $e_5\in C(0;e_1,e_3,e_4)$ 
to define $Q_i^a$, obtaining equivalent phases. Including both 
$e_5$ and $e_6$ gives a 3-parameter GLSM describing the complete 
resolution of the singularity.

The higher order $Y^{pq}$s contain multiple interior points 
corresponding to some or all of the lower order $Y^{pq}$s. Analyzing 
their phase structure using a multiple parameter GLSM exhibits 
phases corresponding to various partial/complete resolutions 
involving lower order $Y^{pq}$ spaces.

Similarly we can see that the higher order $L^{a,b,c}$ spaces typically 
contain blowup modes giving lower order $L^{a,b,c}$s in their partial 
resolutions.


\end{document}